\documentclass[11pt]{article}
\usepackage{graphicx}
\usepackage{amssymb,amsthm,amsmath,float}
\usepackage{epstopdf}
\usepackage{booktabs}  

\usepackage[total={6.5in,8.75in}, top=1.2in, left=0.9in, includefoot]{geometry}

\newcommand{\Thm}[1]{Thm.~\ref{thm:#1}}
\newcommand{\Lem}[1]{Lem.~\ref{lem:#1}}
\newcommand{\Cor}[1]{Cor.~\ref{cor:#1}}
\newcommand{\Sec}[1]{\S \ref{sec:#1}}
\newcommand{\Fig}[1]{Fig.~\ref{fig:#1}}
\newcommand{\Tbl}[1]{Tbl.~\ref{tbl:#1}}
\newcommand{\App}[1]{App.~\ref{app:#1}}

\newcommand{\Eq}[1]{(\ref{eq:#1})}

\newcommand{\InsertFig}[4]
{\begin{figure}[!htb]
      \centerline{
        \includegraphics[width=#4]{./#1}
      }
      \caption{{\footnotesize  #2}
      \label{fig:#3}}
\end{figure}}

\newcommand{\bR}{{\mathbb{ R}}}

\newcommand{\bT}{{\mathbb{ T}}}
\newcommand{\bZ}{{\mathbb{ Z}}}
\newcommand{\bN}{{\mathbb{ N}}}
\newcommand{\bQ}{{\mathbb{ Q}}}

\newcommand{\cD}{{\cal D}}

\newcommand{\cL}{{\cal L}}
\newcommand{\cO}{{\cal O}}

\newcommand{\cR}{{\cal R}}

\newcommand{\eps}{\varepsilon}

\DeclareMathOperator{\sgn}{sgn}
\DeclareMathOperator{\tr}{tr}

\DeclareMathOperator{\diag}{diag}

\DeclareMathOperator{\rank}{rank}
\DeclareMathOperator{\spn}{span}

\DeclareMathOperator{\lcm}{lcm}

\newtheorem{thm}{Theorem}

\newtheorem{cor}[thm]{Corollary}
\newtheorem{lem}[thm]{Lemma}

\theoremstyle{definition}

\newtheorem{rem}{Remark}

\newenvironment{example}{\smallskip\noindent \textbf{Example.}}{\bigskip}

\newcommand{\beq}[1]{\begin{equation}\label{eq:#1}}
\newcommand{\eeq}{\end{equation}}

\newcommand{\bsplit}[1]{\begin{equation}\label{eq:#1}\begin{split}}
\newcommand{\esplit}{\end{split}\end{equation}}

\title{Resonances and Twist in Volume-Preserving Mappings}
\author{  
       H.~R. Dullin
  \\
     School of Mathematics and Statistics\\
     The University of Sydney\\
     Sydney, NSW 2006, Australia\\
     {\tt hdullin@usyd.edu.au} \medskip \\
     and J.~D.~Meiss\thanks
     {
       JDM was supported in part by NSF grant DMS-0707659 and as a visiting 
       research fellow to the University of Sydney. 
       HRD was supported in part by ARC grant DP110102001.
     }\\
    Department of Applied Mathematics\\
    University of Colorado \\
    Boulder, CO 80309-0526, USA\\
   {\tt James.Meiss@colorado.edu} 
}
\date{\today}
\begin{document}
\maketitle

\begin{abstract}
\vspace*{1ex}
\noindent
The phase space of an integrable, volume-preserving map with one action and $d$ angles is foliated by a one-parameter family of $d$-dimensional invariant tori. Perturbations of such a system may lead to chaotic dynamics and transport. We show that near a rank-one, resonant torus these mappings can be reduced to volume-preserving ``standard maps." These have twist only when the image of the frequency map crosses the resonance curve transversely. We show that these maps can be approximated---using averaging theory---by the usual area-preserving twist or nontwist standard maps. The twist condition appropriate for the volume-preserving setting is shown to be distinct from the nondegeneracy condition used in (volume-preserving) KAM theory.
\end{abstract}



\section{Introduction}
Volume preserving maps are appropriate models for many systems including fluid \cite{Cartwright96, Meleshko99, Sotiropoulos01, Rodrigo03, Speetjens04, Anderson06,Mullowney08} and magnetic field line flows \cite{Thyagaraja85, Greene93,Bazzani98} and even the motion of comets perturbed by a planet on an elliptical orbit \cite{Liu94}. One fundamental question relevant for applications is to understand transport in these maps \cite{Piro88, RomKedar93, Cartwright94, Lomeli09b}.

Transport is prohibited between any two regions that are separated by a codimension-one surface. For volume-preserving maps, these dividing surfaces are often tori. Even when there is no dividing surface, we may imagine, as in the two-dimensional case \cite{MacKay84}, that remnants of such surfaces---if they exist---may impede transport.
Similarly, transport may be reduced because of the ``stickiness" of invariant tori, a phenomenon well-established numerically for area-preserving maps \cite{Meiss86, Cristadoro08, Venegeroles09} and for nearly integrable Hamiltonian systems \cite{Poshel93,Perry94}, if not yet completely understood. For three-dimensional maps, stickiness has been observed as algebraic decay of exit time distributions for some systems \cite{Mullowney08} but not for others \cite{Sun08}.

Codimension-one invariant tori are a common feature of the dynamics of three-dimensional volume-preserving mappings.  Indeed these maps could be regarded as ``integrable" when their phase space is foliated by a family of two-dimensional tori on which the dynamics is conjugate to a rigid rotation. We can think of this case as a map with two angles and one action \cite{Piro88}. KAM theory implies that two-tori are robust features of nearly-integrable, one-action maps \cite{Cheng90b, Xia92}. By contrast, in maps with one angle and two actions (which might also be called integrable), the invariant circles are apparently not robust \cite{Mezic01}.

Tori also arise naturally through bifurcation. For example, a fixed point with multipliers $(e^{2\pi i\omega}, e^{-2\pi i\omega}, 1)$ generically undergoes a ``saddle-center-Neimark-Sacker" bifurcation that can create an invariant circle surrounded by a family of two-tori \cite{Dullin09a}. This bifurcation has also been studied in the dissipative case where it can lead to attracting invariant tori as well as strange attractors \cite{Broer08a, Broer08b}. Tori are often destroyed by resonant bifurcations, and this phenomenon is the topic of the current paper.


We begin, in \Sec{OneAction}, with a short discussion of integrable, volume preserving maps with one action and $d$-angles, and the frequency map, $\Omega: \bR \to \bR^d$, that characterizes their dynamics.  Our first goal is to obtain a ``standard" volume-preserving mapping that describes, as Chirikov's standard area-preserving mapping does, the destruction of these invariant tori as the strength of the resonant terms grows. We begin in \Sec{Example} with a numerical study of a simple, three-dimensional example; similar examples have been studied in \cite{GrindrodReardon}. As we will see, this example exhibits many of the features of general one-action maps, near a ``rank-one" resonance.

In \Sec{SingleResonance} we consider more generally a small perturbation of the integrable, one-action map.  Zooming in to a neighborhood of a rank-one resonance gives rise to a map that has a phase space with $d-1$ fast angles, one slow resonant angle and one slow action. The three-dimensional version of this map (when there is one fast angle) is---to first order---equivalent to the example studied in \Sec{Example}.

The main tool for nearly-resonant dynamics is averaging theory. We will show, in \Sec{Averaging}, that volume-preserving normal form theory can be used to transform to new variables that approximately decouple the slow dynamics. This results in an approximate translation symmetry along the fast angles. The normal form can be computed to arbitrary order in the perturbation strength and can be made volume-preserving. The approximate symmetry becomes an ``exact" symmetry if the high-order, fast fluctuations  are discarded, or equivalently if this system is averaged over the fast variables. A symmetry reduction then leads to a map on the slow phase space that is again volume preserving. As is usual for averaging theory, the resulting averaged system is shown to closely approximate the full dynamics, but only on a finite, but long, time scale, see \Thm{Averaging}.

Nevertheless, the qualitative features of the decoupled, averaged map agree surprisingly well with the numerical results of \Sec{Example}, provided that one is far from ``resonance overlap." In particular, when there is one fast angle, the reduced slow map becomes  Chirikov's area-preserving standard map, or its non-twist cousin, depending on properties of the frequency map $\Omega$. The (2D) resonance structure and Chirikov's (2D) overlap criterion agree well with slices of the full three-dimensional dynamics.

Violation of the twist-condition for area-preserving maps leads to meandering curves \cite{HowHoh84,Simo98,Wurm05} and twistless bifurcations \cite{DMS98b}.
We will see in \Sec{Example} and more generally in \Sec{SingleResonance} that the 
transversality conditions whose violation leads to the same dynamical phenomena in higher-dimensional, volume-preserving maps is {\em not} the violation of a Kolmogorov nondegeneracy condition, but instead the condition that the image of $\Omega$ be transverse to the resonance surface in frequency space. Our condition is weaker: KAM theory may still hold at a tangency. This is an essential difference to the symplectic case and shows that  KAM theory ``knows'' little about the qualitative dynamics. 

A similar analysis (blowing-up, decoupling, and averaging)  applies to rank-two resonances, as we discuss in \Sec{RankTwo}, though in this case there are three slow variables. The resulting reduced, three-dimensional, slow map is becomes a quasi-periodic version of the standard map.

\section{One-Action Maps}\label{sec:OneAction}
Though applications of volume-preserving dynamics mainly occur in three-dimensions, where two-dimensional tori are barriers, it is easy to generalize to $d$-tori in $d+1$ dimensions. 
The unperturbed model is the integrable family of maps $f_0: \bT^d \times \bR \to \bT^d \times \bR$ given by
\beq{integrable}
	f_0(x,z) = (x+\Omega(z) \mbox{ mod } 1, \, z) \;.
\eeq
Such a map is called a {\em one-action} map: the variable $z$ plays the role of the action, and the $d$-angles, $x_i\;, i = 1,\ldots d$, are taken to have period one \cite{Cartwright96,Cartwright99, Mezic01}.  This map can be taken to be a model for the behavior in the neighborhood of a smooth, codimension-one invariant torus on which the dynamics is conjugate to rigid rotation.
The {\em frequency map}  (or perhaps, using the terminology from flows, the ``frequency-ratio" map),
\[
	\Omega: \bR \to \bR^d
\]
traces a curve in the frequency space, $\omega = \Omega(z)$, as the ``action" varies, as sketched in \Fig{cantorset}. If we fix a lift for the periodic component of the map $f_0$ then $\Omega(z^*)$ is the rotation number for orbits on the torus $z = z^*$.


\InsertFig{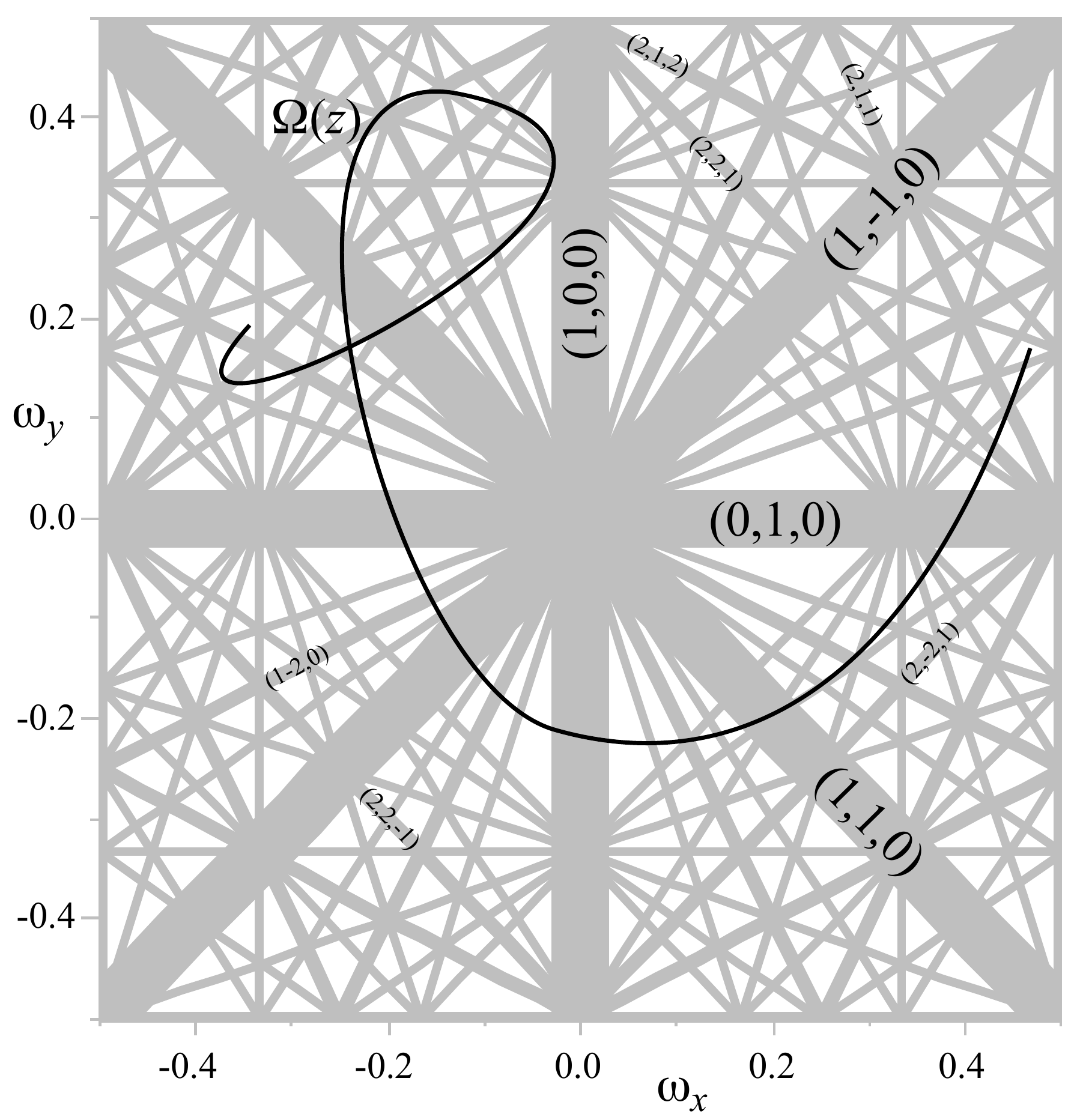}{Resonance web, thickened by the Diophantine condition \Eq{Diophantine} and a frequency map $\Omega$ for a one-action map on $\bR \times \bT^2$. The complement of the web is the positive measure cantor set of Diophantine rotation vectors.}{cantorset}{4in}

A rotation vector $\omega \in \bR^d$ is said to be resonant if there exists a nonzero $(m,n) \in \bZ^d \times \bZ$ such that 
\beq{resonance}
	m \cdot \omega = n \;;
\eeq
if there are no such integers, then $\omega$ is \emph{incommensurate}.
The collection of $d-1$ dimensional planes 
\[
   \cR \equiv \left\{\omega \in \bR^d: m\cdot \omega =n,\;(m,n)\in \bZ^{d+1} \setminus \{0\} 
   			\right\}
\]
is the resonance web; it is a dense subset of $\bR^d$. 

Our goal is to understand the effects of perturbation and resonance on the tori of  \Eq{integrable}. As we will see, it is important to distinguish between resonances of different ranks.
For a given $\omega$, the {\em resonance module} is the sublattice of $\bZ^d$ that corresponds to all integer vectors $m$ that satisfy \Eq{resonance}, namely,
\beq{module}
	\cL(\omega) \equiv \{ m \in \bZ^d: m\cdot\omega \in \bZ \} \;.
\eeq
When $\omega$ is incommensurate $\cL(\omega)$ is trivial: it contains only the origin. When this set is $r$-dimensional, i.e., is the integer span of $r$ independent vectors, then the resonance has {\em rank} $r$. When $\Omega(z^*)$ is $r$-resonant, the invariant $d$-torus $z = z^*$ of \Eq{integrable} consists of families of orbits that are dense on (collections of) $(d-r)$-tori, see
\App{Integrable}.

The $d$-dimensional invariant tori of \Eq{integrable} are examples of rotational tori; generally, a \emph{rotational torus} is a set that is homotopic to the horizontal torus $\{(x,0): x \in \bT^d\}$.\footnote
{Rotational tori need not be graphs over the angles, see \Sec{Example}. 
}
Upon perturbation, many of the rotational invariant tori of \Eq{integrable} will be destroyed. In particular, the resonant tori are fragile---we expect that they will be immediately destroyed and replaced by a finite number of lower dimensional tori and perhaps by families of nonrotational, $d$-dimensional tori. Conversely, KAM theory teaches us that for a torus to be robust it must be not only incommensurate, but \emph{sufficiently} incommensurate---typically it must have a Diophantine frequency vector. The set of Diophantine frequency vectors is
\beq{Diophantine}
	\cD(c,\mu) = \{ \omega \in \bR^d : |m \cdot \omega -n| \ge c|m|^{-\mu}
				\quad \forall (m,n) \in \bZ^{d+1}\setminus\{0\} \} \;,
\eeq
where $|m|$ is any norm, for example the maximum norm.
Even though $\cR$ and $\cD$ are disjoint and $\cR$ is dense in $\bR^d$, $\cD$ is a positive measure Cantor set for each $\mu > d$ and sufficiently small $c > 0$, see \Fig{cantorset}.

KAM theory can indeed be applied to volume-preserving perturbations of the one-action map \Eq{integrable}; it implies, under certain smoothness and nondegeneracy assumptions, that there exists a set of invariant, Diophantine $d$-tori (with $\mu > d^2 + d - 1$) whose measure approaches one as the perturbation goes to zero \cite{Cheng90b, Xia92}.
In these KAM-type theorems a sufficient nondegeneracy condition is that the Wronskian of $D\Omega$ 
must be bounded away from zero.

The curve $\Omega(z)$ will typically intersect planes corresponding to 1-resonances for a dense set of $z$ values.
In this paper we want to emphasize that the dynamical behavior of the perturbed map near these intersections depends in detail on the rank of the resonance and on whether the intersection is transverse or tangent. When the intersection with the resonance plane is transverse, we will show in \Sec{SingleResonance} that the perturbed map can be approximated  by the area-preserving {\em standard map} that drives additional fast angles. By contrast, when the curve $\Omega(z)$ is tangent to the resonance plane, the map can be approximated by the area-preserving {\em standard nontwist map} that drives additional fast angles.

\section{Example: Transverse and Tangent Resonances}\label{sec:Example}

We start by studying the dynamics of an example of a perturbation of $f_0$, a map $(x',z') = f_\eps(x,z)$ given by
\bsplit{theMap}
	x' &= x + \Omega(z') \mod 1 \;, \\
	z' &= z + \eps g(x) \;,
\end{split}\end{equation}
with $g(x+m) = g(x)$ for all $ x \in \bT^d$ and $m \in \bZ^d$.
Here we study the case $d=2$, but in \Sec{SingleResonance} we consider arbitrary $d$ as well as a more general perturbation of \Eq{integrable}.

The map \Eq{theMap} is one natural generalization of the Chirikov standard map (a one-action, one-angle map) and its symplectic cousin, the Froeshl\'e map \cite{Meiss92} (a two-action, two-angle map). It  is volume-preserving with the standard volume form
$
	\wedge dx_1 \wedge \ldots \wedge dx_d \wedge dz
$;
it is not hard to see that the Jacobian, $Df_\eps$, has determinant one.
The map \Eq{theMap} has no rotational invariant tori unless the function $g$ has zero average:
\beq{zeroFlux}
	\int_{\bT^d}  g(x) dx_1 \ldots dx_d = 0 \;.
\eeq
This is precisely the condition that $f_\eps$ is an \emph{exact} volume-preserving map, or equivalently that it has \emph{zero net flux}.
Since we are interested in the persistence and destruction of invariant tori, we will assume that the zero flux condition \Eq{zeroFlux} holds.

For the three-dimensional case, the image of the frequency map is a curve in $\bR^2$ that can locally be approximated by a parabola. As we will see more generally in \Sec{SingleResonance}, coordinates can be chosen near a rank-one resonance so that the frequency map becomes
\beq{Parabola}
	\Omega(z) = (\gamma + z, -\delta + \beta z^2) \;.
\eeq
This parabola is sketched in \Fig{resonances}.
The twist condition of the volume-preserving KAM-type theories in this case becomes $\beta \neq 0$. 

\InsertFig{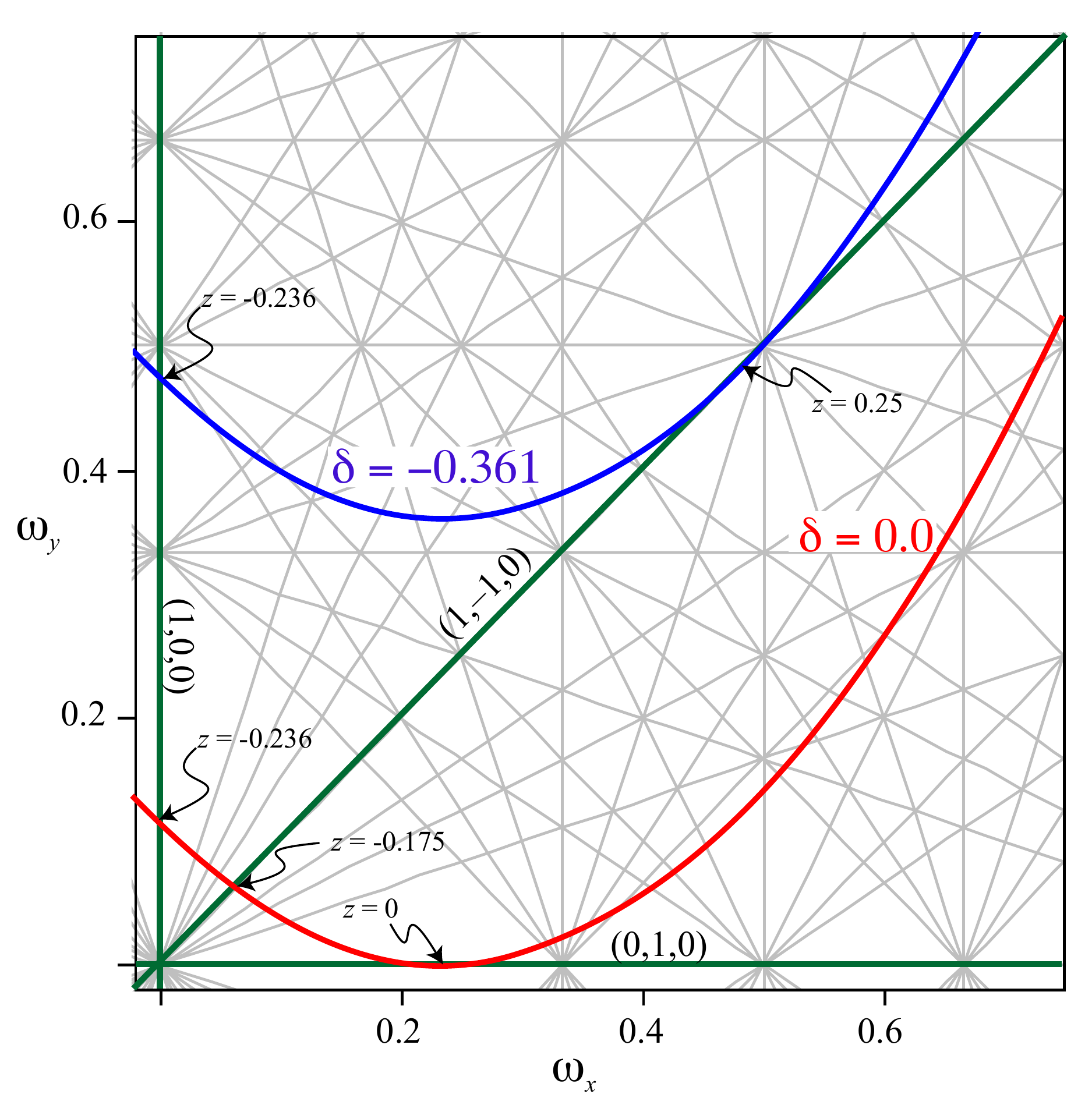}{Resonance lines $m\cdot\omega = n$ for $|m| \le 3$,
and the image of the frequency map \Eq{Parabola} for $\beta = 2$, $\gamma = \sqrt{5}-2 \approx 0.236$, and two values of $\delta$ that give rise to tangencies of $\Omega$ with the forced resonances of \Eq{Froeshle}.}{resonances}{4in}

Using the frequency map \Eq{Parabola}, \Eq{theMap} takes the ``standard" form
\bsplit{stdTangency}
	x' &= x + z' + \gamma \;, \\
	y' &= y - \delta +  \beta z'^2 \;, \\
	z' &= z + \eps g(x,y) \;.
\end{split}\end{equation}
The forcing function $g(x)$ in \Eq{stdTangency} is periodic; therefore it can be expanded in a Fourier series. Since $g(x)$ has zero mean \Eq{zeroFlux}, its $(0,0)$ Fourier component must vanish. In this section, for simplicity and to illustrate the general theory of \Sec{SingleResonance}, we choose $g$ to have just three terms:
\beq{Froeshle}
	g(x,y) = -a\sin(2\pi x) - b \sin(2 \pi y) - c \sin(2\pi(x-y)) \;.
\eeq
The terms with amplitudes $a,b,c$ represent forcing at the resonances $(1,0,n)$, $(0,1,n)$, and $(1,-1,n)$, respectively. These forced resonances, $m \cdot \Omega(z^*) = n$, occur at positions $z^*$ listed in \Tbl{resonances}. 

\begin{table}[bt]
   \centering
   \begin{tabular}{@{} ccccc @{}} 
      \toprule 
       $(m,n)$  &  Amplitude & $z^*$ & $M$ & $\delta_T$ \\
      \midrule
      $(1,0,n)$	& $a$ & $n-\gamma$ &  $1$ & none \\ 
      $(0,1,n)$	& $b$ & $\pm \sqrt{\frac{\delta+n}{\beta}}$ 
      			    & $\frac{1}{2\beta z^*}$
				    & $-n$ \\
      $(1,-1,n)$	& $c$ & $\frac{1}{2\beta}
      				\left[1 \pm\sqrt{1+4\beta(\delta+\gamma-n)}\right]$
					& $\frac{1}{1-2\beta z^*}$
					& $n-\frac{1}{4\beta} -\gamma$ 
					\\ 
      \bottomrule
   \end{tabular}
   \caption{Forced resonances. $m \cdot \Omega(z^*) = n$ for \Eq{stdTangency}, their effective masses $M$, and values of $\delta$ for a tangency.}
   \label{tbl:resonances} 
\end{table}

Since $\Omega(z)$ is a parabola, it may intersect a resonance line $m \cdot \omega = n$ more than once; however generically, it will cross transversely. When $\eps \ll 1$ the dynamics near a transverse crossing of a rank-one resonance can be analyzed by expanding the map near $z = z^*$ and performing an appropriate average over the nonresonant, fast angle using the averaging theorem presented in \Sec{Averaging}.

For example, the $(1,0,0)$ resonance occurs when $\Omega_x(z_a) = \gamma + z_a = 0$. Generically the second frequency $\Omega_y(z_a) = \beta\gamma^2 -\delta \equiv \omega_0$ will be irrational so that the resonance  has rank one with resonance module
\[
	\cL((0,\omega_0)) = \{(k,0): k \in \bZ\} \;, \quad \omega_0 \not\in \bQ \;.
\]
To expand near this resonance, define new variables\footnote
{We switch the order of $x$ and $y$ to agree with the general notation of \Sec{Averaging}}
\[
	(\xi,\eta,\zeta) = \big(y,x,\eps^{-1/2}(z - z_a)\big) \;.
\]
In these coordinates, \Eq{stdTangency} reduces to
\beq{StandardZoomed}
\begin{aligned}
	\xi' &= \xi + \omega_0  + \cO(\sqrt{\eps}) \;, \\
	\eta' &= \eta + \sqrt{\eps} \zeta' \;,\\
	\zeta' &= \zeta + \sqrt{\eps} g(\eta,\xi) \;.
\end{aligned}
\eeq
When $\eps \ll 1$, the angle $\xi$ is rapidly rotating relative to $\eta$,
so that the forcing terms in \Eq{Froeshle} proportional to $b$ and $c$  rapidly fluctuate relative to the term proportional to $a$. Thus one would expect that the dynamics of the slow variables $(\eta, \zeta)$ could be approximated by simply averaging over the fast angle, giving the two-dimensional area-preserving map
\bsplit{stdMap}
	 \eta' &=   \eta + \sqrt{\eps}  \zeta'  \;, \\
	 \zeta' &=  \zeta - \sqrt{\eps}a \sin(2\pi  \eta) \;.
\end{split}\eeq
Indeed, this is verified in \Sec{Averaging} provided that the fast frequency, $\omega_0$, is Diophantine \Eq{Diophantine}. The transformations of \Sec{Averaging}---to first order---then give the map \Eq{stdMap}. Theorem \ref{thm:Averaging} implies that an orbit of \Eq{stdMap} is $\cO(\eps)$-conjugate to the projection of an orbit of the original map onto the slow variables $(\eta,\zeta)$ on the time scale $ t \sim \cO(\eps^{-1/2})$.  This is valid so long as the orbit of \Eq{stdMap} stays in an $\cO(1)$ neighborhood of $\zeta = 0$.
The quality of approximation can be increased to $\cO(\eps^{n/2})$ for arbitrary $n$ by 
the iterative procedure described in \Sec{Averaging}.

Note that \Eq{stdMap} is symplectic. Indeed, when $\eps \ll 1$, it is approximately the time $\sqrt{\eps}$ map of the one-degree-of-freedom pendulum Hamiltonian
\beq{Pendulum}
	H(\eta,\zeta; M, a) = \frac{\zeta^2}{2M} - \frac{a}{2\pi} \cos(2\pi \eta) \;,
\eeq
with momentum $\zeta$, coordinate $\eta$, an effective ``mass" $M = M_a = 1$ and amplitude $a$.

Consequently, in a $\sqrt{\eps}$ neighborhood of $z = -\gamma$, the orbits lie on contours of constant $H$ to $\cO(\eps)$ for times $\cO(\eps^{-\frac12})$. In the original three-dimensional map, the pendulum's librating orbits correspond to tubes aligned with the $y$-axis; two such orbits are shown in \Fig{TransverseResonances} (the red and purple orbits). The rotating orbits of \Eq{Pendulum} correspond to rotational invariant tori of \Eq{stdTangency}. The full width of this resonance in $z$ is the width of the separatrix contour of $H$, appropriately scaled in $\eps$,
\beq{Width}
	W(M,a) = \sqrt{\tfrac{8\eps}{\pi}|M a|} \;.
\eeq
When $\eps a = 0.005$ as in \Fig{TransverseResonances}, this gives $w_a = W(1,a) \approx 0.11$.

\InsertFig{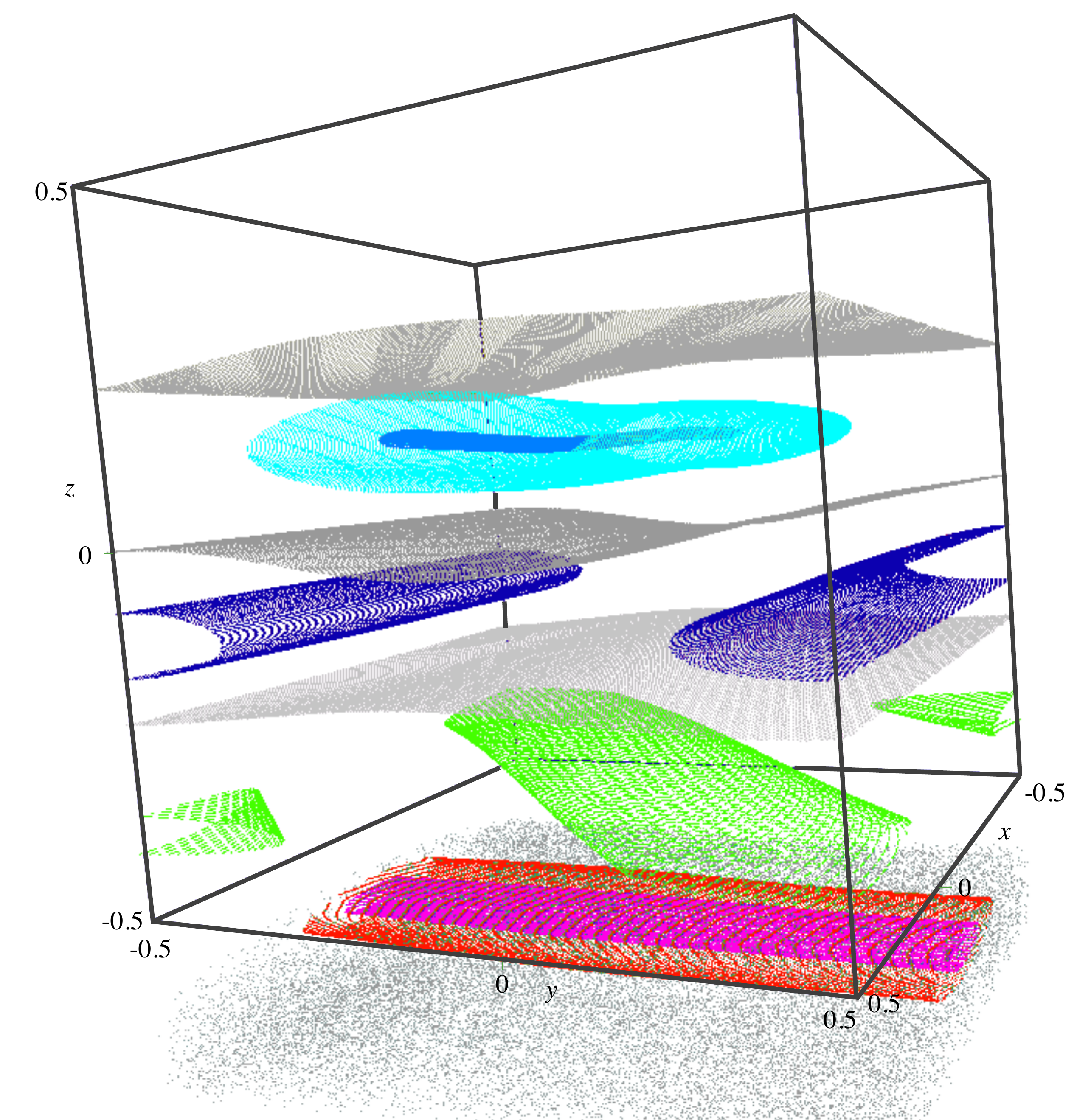}{Phase space of \Eq{stdTangency} for $\beta = 2$, $\gamma = \tfrac12(\sqrt{5}-1) \approx 0.618$, $\delta = 2\delta_R \approx 0.0284$, $a=b=c=1$, and $\eps = 0.005$. Orbits trapped in four resonances are shown. The three blue-toned orbits correspond to the two $(0,1,0)$ resonances at $z_b^\pm = \pm 0.119$. The green orbit is trapped in the $(1,-1,0)$ resonance at $z_c^-=-0.371$, and the two red-toned orbits librate in the $(1,0,0)$ resonance at $z_a = -\gamma = -0.618$. Also shown are three rotational tori and a chaotic orbit (grey). }{TransverseResonances}{4in}

A similar analysis can be done for the $(0,1,0)$ resonance.
If $\delta\beta > 0$, the frequency map \Eq{Parabola} crosses this resonance at two 
positions, $z_b^\pm$, given in \Tbl{resonances}. This resonance has rank one when $
\omega_0 = \Omega_x(z_b) = \gamma + z_b \not\in \bQ$, in which case the same expansion 
procedure can be applied for the variables $(\xi,\eta,\zeta) = (x,y,\eps^{-1/2}(z-z_b))$, 
since the fast angle is now $x$. If the fast frequency is Diophantine, \Thm{Averaging} 
applies and the averaged Hamiltonian is again \Eq{Pendulum} with new arguments: $H(y,
\zeta; M_b,b)$. The effective mass $M_b$, given in \Tbl{resonances}, is real precisely 
when the resonance crossing is transverse: $\delta \beta > 0$. In this case the resonance 
width is $w_b = W(M_b,b)$ with the same function \Eq{Width}. Since the effective masses of 
the $z_b^\pm$ resonances have opposite signs, the resonance with $M_b b > 0$ is an island 
centered at $y = 0$, while that with $M_b b < 0$ is phase shifted, and is centered at $y = 
\tfrac12$. The three blue-toned orbits shown in \Fig{TransverseResonances} are trapped in 
these resonances. For the parameters of this figure the resonances are centered at $z_b^
\pm = \pm 0.119$ and have widths $w_b \approx 0.16$. Since the separation between the 
resonances exceeds the sum of their half-widths, the Chirikov overlap criterion leads to 
the expectation that there are rotational invariant tori between the two resonances \cite
{Lichtenberg92}. Indeed, this is what we observe numerically; one such separating torus 
(in grey) is shown in \Fig{TransverseResonances}.

The frequency curve \Eq{Parabola} also crosses the $(1,-1,0)$ resonance transversely at the two points $z=z_c^\pm$ given in \Tbl{resonances} provided that $\beta(\delta + \gamma) > -\tfrac14$. This resonance can be treated as before upon defining new  variables 
\beq{110transformation}
	(\xi,\eta,\zeta) = (y,x-y, \eps^{-1/2}(z-z_c)) \;,
\eeq
aligned with the resonance. The map \Eq{stdTangency} now becomes
\begin{align*}
	\xi' &= \xi + \omega_0 + \cO(\sqrt{\eps}) \;, \\
	\eta' &= \eta + \sqrt{\eps}\frac{\zeta'}{M_c} \;, \\
	\zeta' &= \zeta +\sqrt{\eps}g(\xi+\eta,\xi) \;,
\end{align*}
with  $\omega_0 = \gamma + z_c$ and $M_c$ given in \Tbl{resonances}. When  $\omega_0 \in 
\cD(c,\mu)$, averaging over $\xi$, according to \Thm{Averaging}, now eliminates the terms 
proportional to $a$ and $b$ in $g$. The resulting system is again  approximated by the 
time $\sqrt{\eps}$ map of \Eq{Pendulum}: $H(\eta,\zeta; M_c,c)$. In \Fig
{TransverseResonances} only the resonance at $z_c^- \approx -0.371$ is visible; it has a 
width $w_c^-\approx 0.072$. 

In the phase portrait of \Fig{TransverseResonances} the resonances are all separated by rotational invariant tori. Indeed the Chirikov overlap parameters between neighboring, forced resonances,\begin{align*}
	s_{b^+b^-} &=  \frac{w_b^+ + w_b^-}{2|z_b^+- z_b^-|} \approx 0.66 \;,\\
	s_{b^-c^-} &= \frac{w_b^- + w_c^-}{2|z_b^- -z_c^-|} \approx 0.47 \;, \\
	s_{c^-a} &= \frac{w_c^- + w_a}{2|z_c^- - z_a|} \approx 0.37 \;,
\end{align*}
are less than the phenomenological threshold value $s = \tfrac23$ for ``global chaos" \cite{Lichtenberg92,Meiss07a}, so we expect---and observe---that rotational invariant tori separate these resonances. Of course, there are also many chaotic trajectories for the parameters of \Fig{TransverseResonances}. For example, the $(1,0,0)$ resonance is surrounded by a large chaotic layer (grey in the figure) because of overlap with a $(1,-1,-1)$ resonance at $z^*= -0.691$ that is not shown in the figure. This resonance and the $(1,0,0)$ resonance have an overlap parameter $s = 1.2$, and are not separated by any rotational tori.

As $\delta$ is varied the resonance locations and widths change as shown in \Fig{ResonanceWidths}. The averaged Hamiltonian description provides a qualitative description near each transversal crossing provided that the resonance widths are small enough to avoid overlap. Of course, even in this case the region near the separatrix of the averaged Hamiltonian is replaced by a chaotic zone for the full map. As we will see in \Sec{SingleResonance}, a similar structure governs the dynamics near any rank-one resonance whenever $\Omega(z)$ crosses the resonance transversely. 

\InsertFig{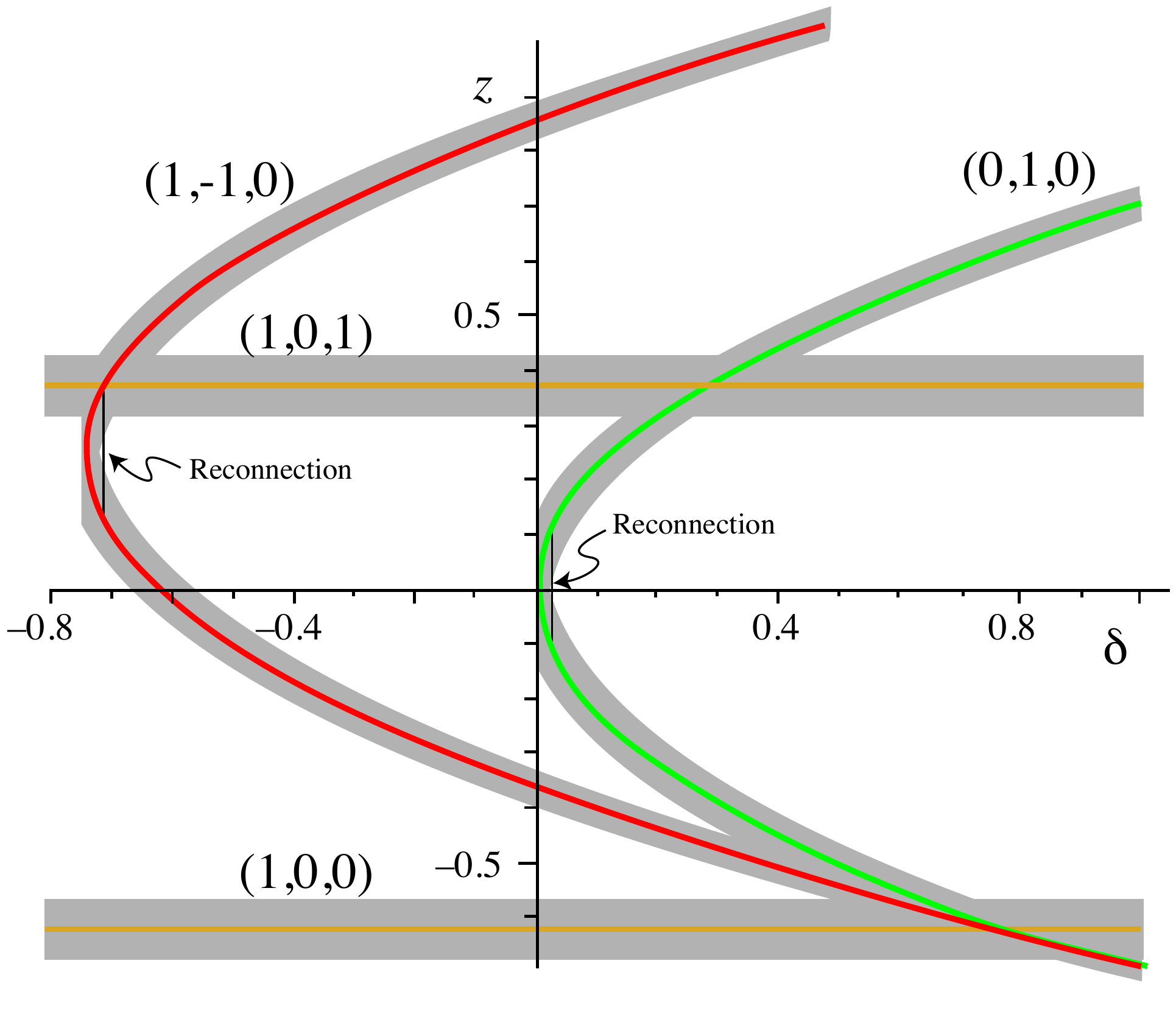}{Resonance centers $z^*$ (curves) and widths (grey regions) as a function of $\delta$, for the parameters of \Fig{TransverseResonances}.}{ResonanceWidths}{4in}

However, when the resonance intersection is nontransversal, the dynamics is radically different. The curves $z^*(\delta)$ for the $(1,0,0)$ and $(1,-1,0)$ resonances in \Fig{ResonanceWidths} are parabolas with turning points, $z^*(\delta_T) = z_T$ that signal nontransversal crossings. Tangency with an $(m,n)$ resonance occurs when
\beq{tangency}
	m \cdot \Omega(z_T) = n  \mbox{ and } m \cdot D\Omega(z_T) = 0 \;. 
\eeq
Since a frequency curve $\Omega(z)$ will generically intersect a resonance line for some $z^*$, tangency is a codimension-one phenomena---it will generically occur if the frequency map $\Omega$ depends upon a single parameter. For the example \Eq{Parabola} we can think of $\delta$ as the parameter that unfolds these tangencies. For the model \Eq{stdTangency}, the $(1,0,n)$ resonance is never tangent but tangencies do occur for both the $(0,1,n)$ and $(1,-1,n)$ resonances where their effective masses diverge. The values of $\delta$ for these tangencies are given in \Tbl{resonances}. 

The frequency curve is tangent to the $(0,1,0)$ resonance when $\delta= \delta_T= 0$ at $z = z_T = 0$. If we assume that $\Omega_x(z_T) = \gamma \in \cD(c,\mu)$,
then the angle $x$ rapidly rotates, and for $z \approx z_T$ and $\delta \approx \delta_T$, the map \Eq{stdTangency} is approximated by the averaged map
\bsplit{Nontwist}
	y' &= y - \delta +  \beta z'^2  \;, \\
	z' &= z - \eps b \sin(2 \pi y)  \;.
\end{split}\end{equation}
Again the averaged system is a two-dimensional, symplectic map.
This system  should approximate the original dynamics when $z$ and $\delta$ are near $z_T = \delta_T =0$. When $\beta = \cO(1)$, a maximal balance of the terms in this map occurs when
\bsplit{scaling}
	z &=  z_T + \eps^{\frac13} \zeta \;, \\
	\delta &=  \delta_T + \eps^{\frac23} \Delta \;.
\end{split}\end{equation}
In this case, \Eq{Nontwist} is approximately the time $\eps^{\frac23}$ map of the Hamiltonian
\beq{HReconnect}
	H_R(y,\zeta;\beta,\Delta,b) = \frac{\beta}{3} \zeta^3 -\Delta \zeta- \frac{b}{2\pi} \cos(2\pi y) \;.
\eeq
When $\Delta\beta > 0$  this Hamiltonian has four equilibria, at $y = 0, \frac12$ and $\zeta = \pm \sqrt{\frac{\Delta}{\beta}}$.  When $\beta b > 0$ the two equilibria
$(0, -\sqrt{\frac{\Delta}{\beta}})$ and $(\frac12,\sqrt{\frac{\Delta}{\beta}})$ are saddles, and the remaining two are centers, see \Fig{Reconnection}. For $|\Delta| > |\Delta_R|$  where
\[
 	 \Delta_R^3 \equiv \frac{9}{16\pi^2}\beta b^2  \;,
\]
there are two independent island chains. The separatrices of these islands correspond to 
the energies $E_{sx} =  \pm \frac{2}{3\sqrt{|\beta|}}( \Delta_R^{\frac32}-\Delta^
{\frac32})$. These coincide at $\Delta = \Delta_R$, where there is a ``reconnection 
bifurcation", see \Fig{Reconnection}. Note that reconnection occurs when $\sgn(\Delta_R) = 
\sgn(\beta)$ at an unscaled value $\delta_R = \delta_T + \eps^{\frac23}\Delta_R$. The 
elliptic and saddle equilibria on the lines $y=0$ and $y=\frac12$ move together as 
$\Delta$ decreases until they are destroyed in saddle-center bifurcations at $\Delta = 0$.

\InsertFig{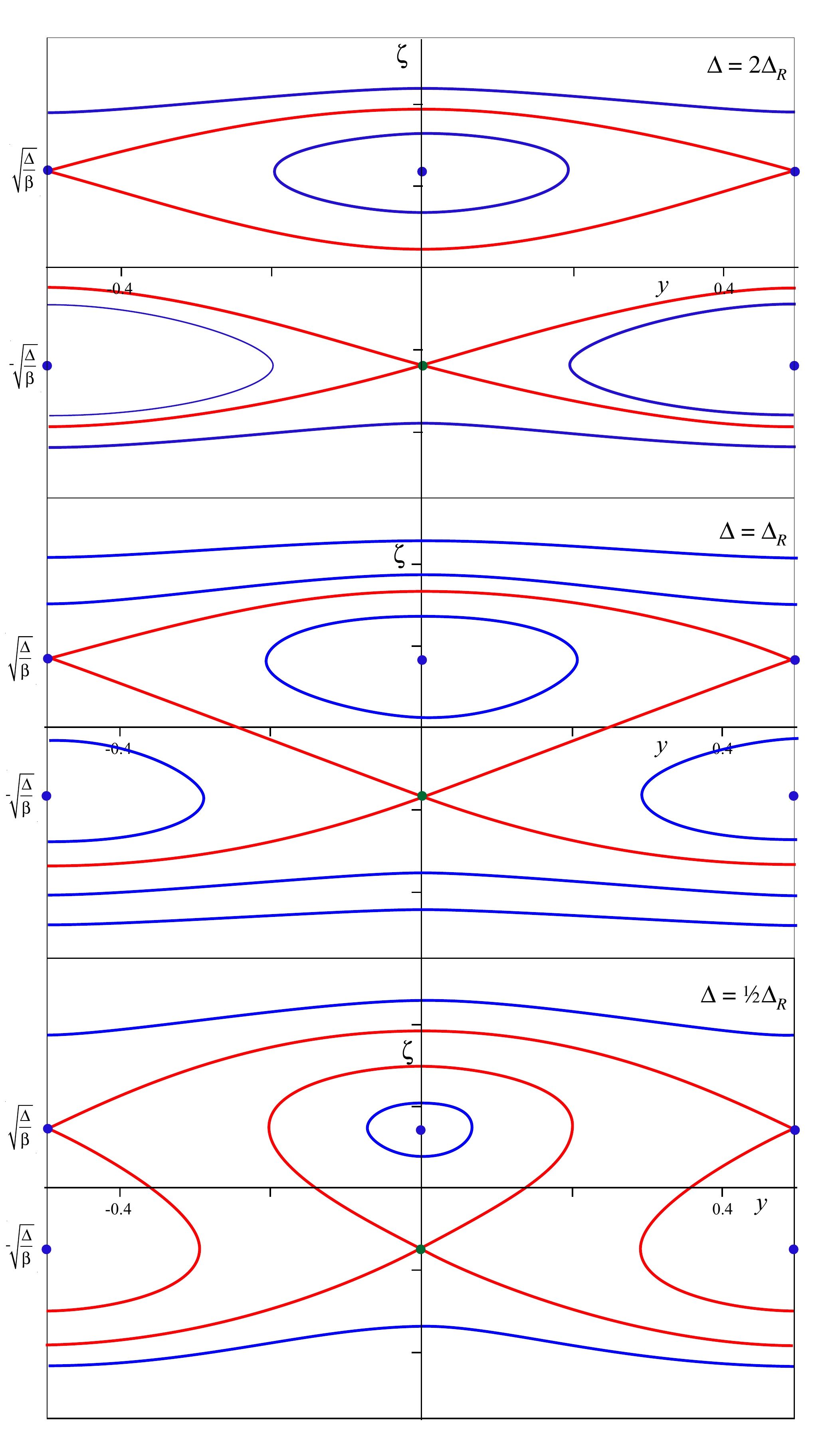}{Contours of the Hamiltonian \Eq{HReconnect} for three values of  $\Delta$ when $\beta b > 0$.}{Reconnection}{3in}

This structure can be most easily seen in the three-dimensional map by using slices. A 
slice is an approximate, two-dimensional phase portrait obtained by plotting orbits only 
when they land within a thin slab. In \Fig{2DeltaR}, we show two such slices, $|x|< 0.001$ 
on the left and $|y|< 0.001$ on the right---thus on average to obtain one point in either 
slice the map must be iterated $\approx 500$ times. These slices show the same orbits. The 
fast dynamics of the $(0,1,n)$ and $(1,-1,n)$ resonances are transverse to the left slice, 
and so the islands described by the slow dynamics can be seen. In particular the two $(0,1,0)$ resonances at $z_b^\pm$ are well described by the upper portrait 
of \Fig{Reconnection} since $\delta_T = 0$, and $\delta = 2\delta_R$ at these parameter 
values. There are also three $(1,-1,n)$ resonances indicated in the figure 
with their widths computed from \Eq{Width}. The $(1,0,n)$ resonances do not appear in this 
slice since their librating orbits move approximately parallel to the slice; however, the 
chaotic orbit above the $(1,-1,-1)$ resonance is due to its overlap with $(1,0,0)$.
Two $(1,0,n)$ island do appear in the right slice; the diagonal $(1,-1,n)$ resonances also 
appear again.
 
\InsertFig{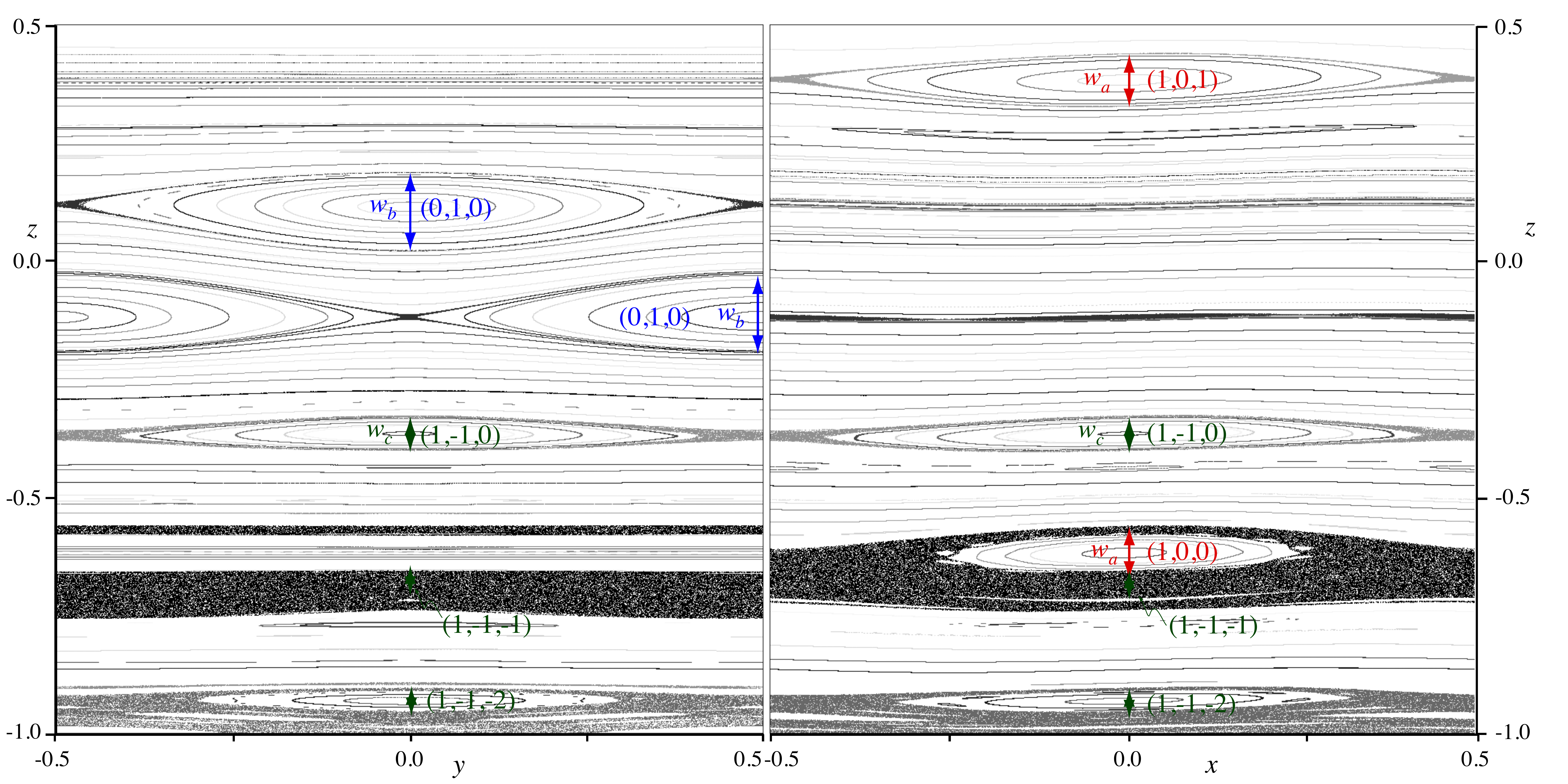}{Two dimensional slices $|x|< 0.001$ (left pane) and $|y| < 0.001$ of the map \Eq{stdTangency} for the parameters of \Fig{TransverseResonances}. The arrows indicate the widths computed using \Eq{Width}. The $x$-slice of the $(0,1,0)$ resonances resembles the top pane of \Fig{Reconnection}.}{2DeltaR}{6.5in}

Similar pairs of slices of the three-dimensional map are shown in \Fig{DeltaR} and \Fig{DeltaRo2} for $\delta = \delta_R$ and $\tfrac12 \delta_R$, respectively. Again, the averaged Hamiltonian portraits from \Fig{Reconnection} are closely mimicked near the $(0,1,0)$ resonance in these figures except that the separatrices become chaotic. In particular, note that the invariant tori trapped between the $z_b^\pm$ resonances in \Fig{DeltaRo2} are ``meandering"---they are not graphs over the unperturbed angles.

\InsertFig{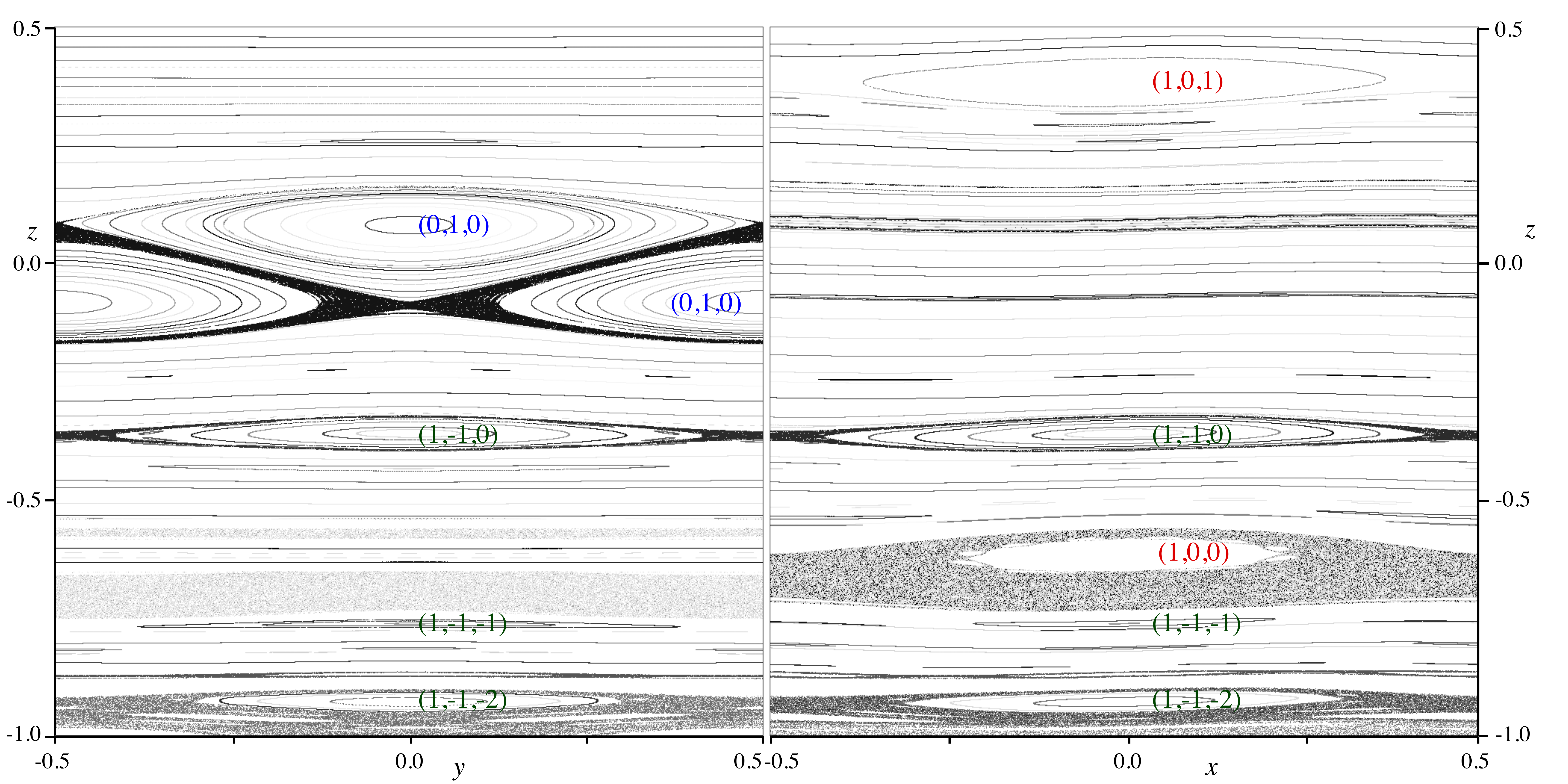}{Two dimensional slices  of the map \Eq{stdTangency} with the parameters of \Fig{2DeltaR}, except $\delta = \delta_R \approx 0.0142$.  The $x$-slice of the $(0,1,0)$ resonances resembles the middle pane of \Fig{Reconnection}.}{DeltaR}{6.5in}
\InsertFig{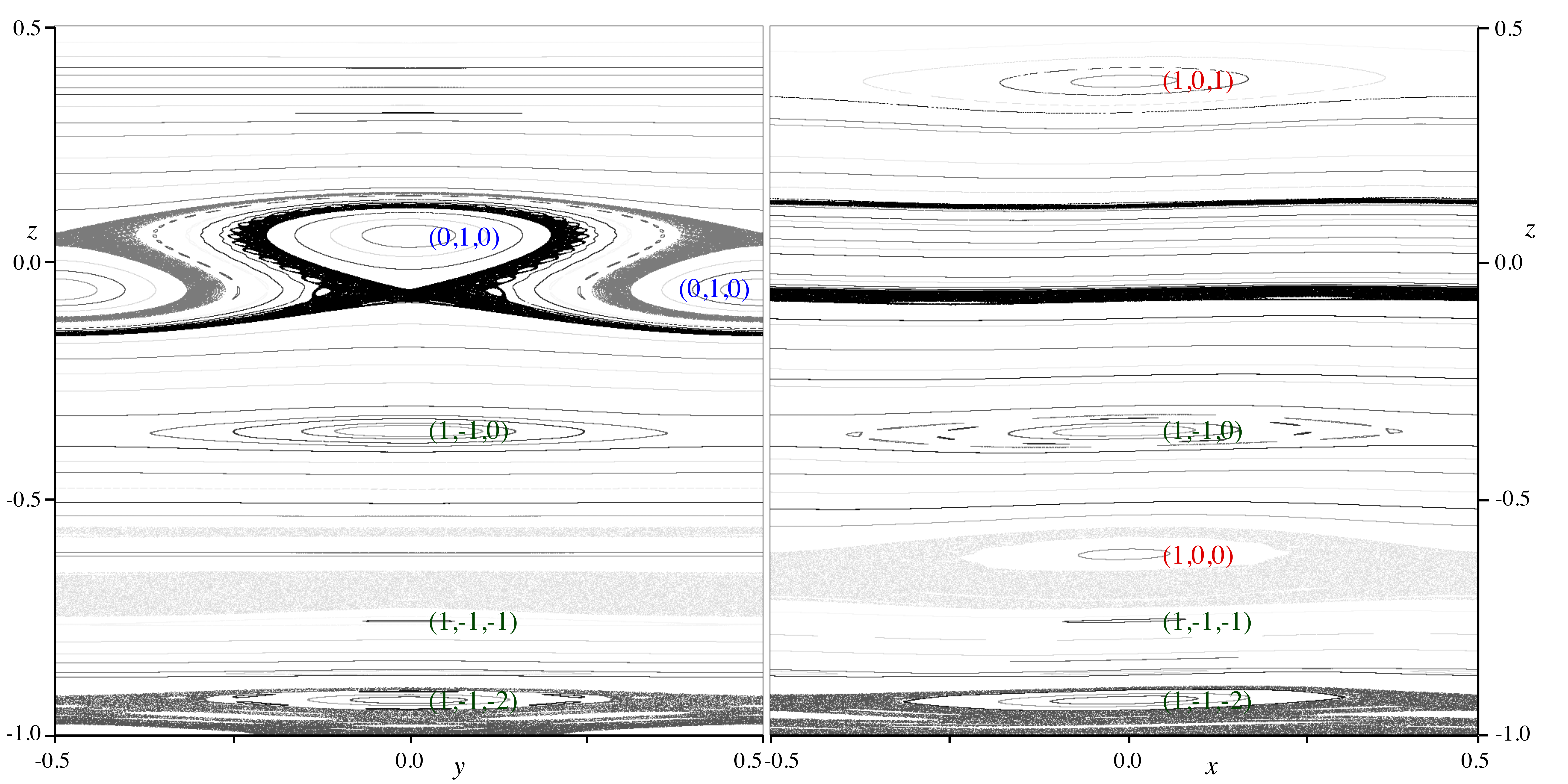}{Two dimensional slices of the map \Eq{stdTangency} with the parameters of \Fig{2DeltaR}, except $\delta = \tfrac12\Delta_R \approx 0.0071$.  The $x$-slice of the $(0,1,0)$ resonances resembles the bottom pane of \Fig{Reconnection}.}{DeltaRo2}{6.5in}

Tangency similarly occurs for the $(1,-1,0)$ resonance when $\delta = \delta_T$ as given in \Tbl{resonances} when $z = z_T = \frac{1}{2\beta}$, where the effective mass $M_c$ is infinite. As before, we must use the aligned angles $(\xi,\eta)$  of \Eq{110transformation} to construct the averaged map, now scaling $z$ and $\delta$ as in \Eq{scaling}. The result is exactly \Eq{HReconnect} with parameters $H_R(\eta,\zeta,-\beta,-\Delta,c)$. Thus the same reconnection scenario applies to this resonance near $4\beta(\delta+\gamma) = -1$. A three-dimensional phase portrait near the tangency is shown in \Fig{110Tangency}. For these parameters, $\delta_T \approx -0.743$, and $\delta_R \approx -0.729$. The figure corresponds to $\Delta = 0.21\Delta_R$. The chaotic layer around the $(1,-1,0)$ resonances overlaps with that around the $(1,0,1)$ resonance, as indicated in \Fig{ResonanceWidths}.

\InsertFig{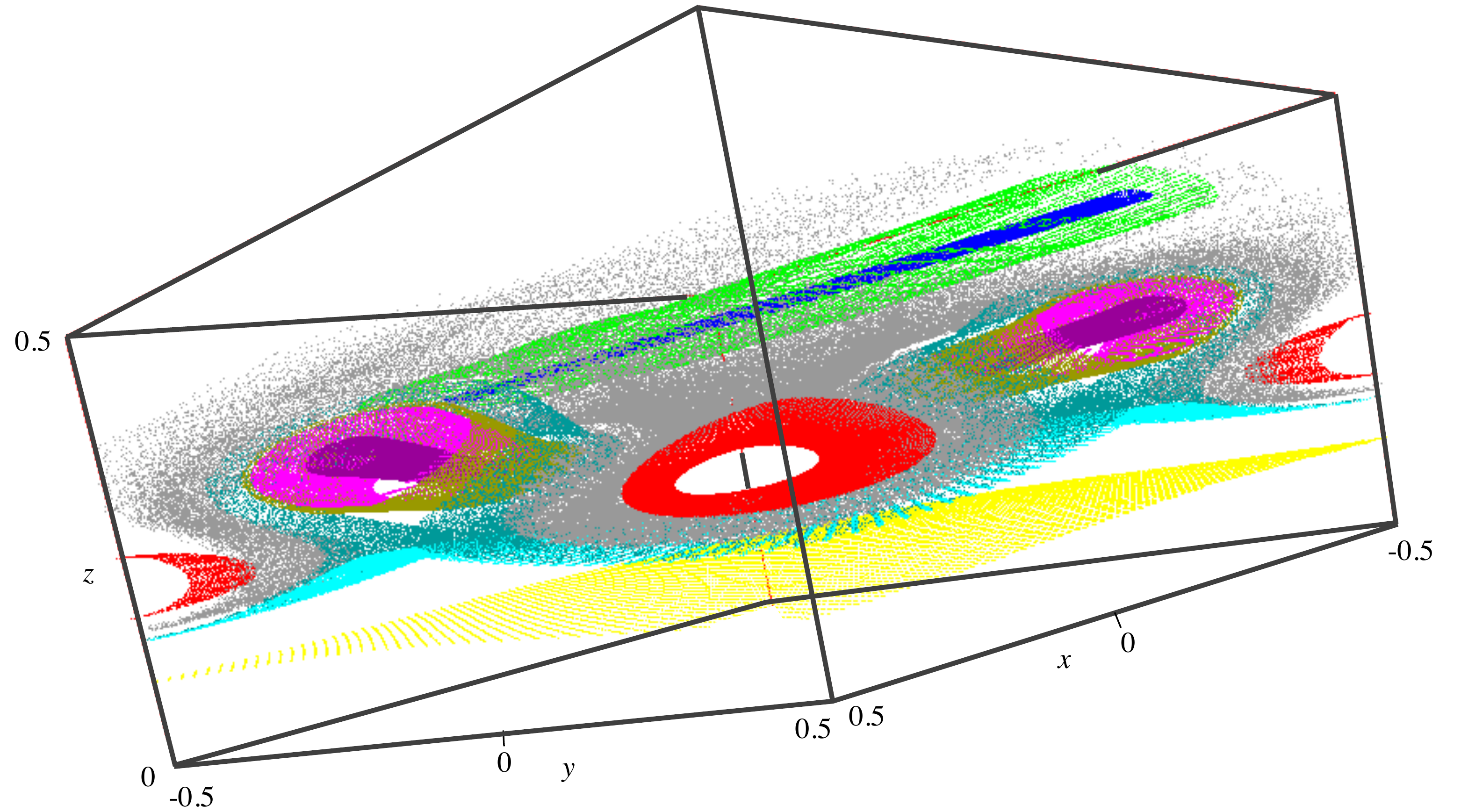}{Orbits of the map \Eq{stdTangency} for $\beta = 2$, $\gamma = \tfrac12(\sqrt{5}-1)$, $\delta = -0.74$, $\eps = 0.001$, $a = 4$ and $b = c = 1$. Shown are orbits in the two, nearly tangent $(1,-1,0)$ resonances (shades of red and light blue) and in the nearby $(1,0,1)$ resonance (green and dark blue). A chaotic orbit (grey) wanders among all three resonances and an invariant torus (yellow) bounds the trajectories below.}{110Tangency}{5in}

\clearpage
	
\section{The General Rank-One Resonance}\label{sec:SingleResonance}
In this section we consider more generally the case of a rank-one resonance of a volume-preserving map $f_\eps: \bT^d \times \bR \to \bT^d \times \bR$ that is an $\eps \ll 1$ perturbation  of the integrable map \Eq{integrable}:
\bsplit{Perturbed}
	x' &= x + \Omega(z') + \eps X(x,z;\eps) \;, \\
	z' &= z + \eps Z(x,z;\eps) \;.
\end{split}\end{equation}
We suppose that \Eq{Perturbed} is volume preserving, 
so that its Jacobian has determinant $|Df_\eps|  = 1$. 
In order to apply the averaging theory (see \Thm{Averaging}), we will assume 
that $X$, $Z$, and $\Omega$ are analytic functions of all their arguments. 
We will also need a Diophantine condition on the fast frequency vector 
at resonance, see below. 

When a point $z^*$ corresponds to a rank-one resonance, the frequency vector $\Omega^* = \Omega(z^*)$ obeys a resonance condition
\[
	m \cdot \Omega^* = n
\]
for exactly one coprime vector $(m,n) \in \bZ^d \times \bZ \setminus \{0\}$. Our goal is to show that \Eq{Perturbed} reduces essentially to \Eq{stdTangency} (with a general function $g$) near $z = z^*$.  

As we discussed in \Sec{Example}, the frequency curve will generically intersect a resonant plane transversely. However, under variation of a parameter, the intersection may become tangent; we assume that any tangency is quadratic. These two cases can be treated simultaneously by expanding about the resonant torus, setting $z = z^* + \cO(\eps^p)$, and choosing the exponent $p$ to be $\tfrac12$ or $\tfrac13$, respectively. In order to unfold the codimension-one tangency, it is convenient to expand about a torus slightly shifted away from the resonance. 
Thus instead of expanding about $z^*$, we expand about a shifted point $\bar z$,
\beq{zScale}
	z = \bar z + \eps^p \zeta \;,\quad 0 < p \le \tfrac12 \;.
\eeq
We also assume that $\bar z = z^* +\cO(\eps^{2p})$ to define the frequency shift $\delta$ by
\beq{freqShift}
	m \cdot \Omega(\bar z) = n - \eps^{2p} \delta \;.
\eeq
Substituting these expansions into \Eq{Perturbed} gives
\bsplit{Expansion}
	x' &= x + \Omega(\bar z) + \eps^p D\Omega(z^*) \zeta' + 
		\tfrac12 \eps^{2p} D^2\Omega(z^*) \zeta'^2 + \cO(\eps,\eps^{3p})	\;,	\\
	\zeta' &= \zeta + \eps^{1-p} Z(x, z^*;0) + \cO(\eps) \;.
\end{split}\end{equation}

Though the components of the vector $(m,n)$ are coprime by assumption, the components of $m$ need not be coprime, see \App{Integrable}. Let $k =gcd(m)$ be their greatest common divisor so that $\hat m = m /k$ is a coprime, integer vector. As discussed in  \App{Integrable}, there exists a basis for the lattice $\bZ^d$ with $\hat m$ as one basis vector; equivalently, there exists a matrix $M \in SL(d,\bZ)$ whose last row is $\hat m^T$. Thus we can define new coordinates $(\xi,\eta)\in \bT^{d-1} \times \bT$ as
\[
	\begin{pmatrix} \xi \\ \eta \end{pmatrix} \equiv M x 
	    \equiv \begin{pmatrix} \hat M x \\ \hat m^T x \end{pmatrix}
\]
to align the last angle $\eta$ with the resonance. 

Since $Z(x,z;\eps)$ is analytic, it has a Fourier series with analytic coefficients $\tilde Z_j(z;\eps)$. Consequently the perturbation $Z(x,z^*;0)$ can be written in terms of the new angles as
\begin{align*}
	g(\xi,\eta) &\equiv Z(M^{-1}(\xi^T,\eta)^T, z^*;0) 
	  = \sum_{j \in \bZ^d} \tilde{Z}_j e^{2\pi i (\xi^T,\eta) M^{-T}j} 
      = \sum_{l \in \bZ^d} \tilde{g}_l \, e^{2\pi i l  \cdot (\xi^T,\eta)} \;,
\end{align*}
so that $g$ has the Fourier coefficients $\tilde{g}_l = \tilde{Z}_{M^T l}$, $l \in \bZ^d$. 

Since $\bar z = z^* + \cO(\eps^{2p})$ by \Eq{freqShift}, we obtain
\[
	k M \Omega(\bar z) =  \begin{pmatrix} 
	                          \hat \Omega^*  \\ n \end{pmatrix} 
				 -\eps^{2p}\begin{pmatrix} \cO(1) \\ \delta \end{pmatrix}\;,
\]
where $\hat \Omega^* = k  \hat M \Omega^*$. In the new coordinates, the map \Eq{Expansion} becomes
\bsplit{ExpandedMap}
	\xi'   &= \xi + \frac{1}{k} \left( \hat{\Omega}^* + \eps^p \lambda \zeta' \right) + \cO(\eps,\eps^{2p}) \;,\\
	\eta'     &= \eta + \frac{1}{k}\left(n  + \eps^p \alpha \zeta' 
			+ \eps ^{2p}(\beta \zeta'^2 - \delta) \right) +\cO(\eps,\eps^{3p}) \;,\\
	\zeta' &= \zeta + \eps^{1-p} g(\eta,\xi) + \cO(\eps)  \;,
\end{split}
\eeq
where
\beq{alphabeta}
	\lambda \equiv k \hat M D\Omega(z^*) \;,  \quad
	\alpha \equiv  m \cdot D\Omega(z^*) \;, \quad 
	\beta \equiv \frac12 m \cdot D^2\Omega(z^*) \;.
\eeq

Iterating this map $k$ times to $\cO(\eps)$ will eliminate the rational $\tfrac{n}{k}$ in the $\eta$ equation, since this angle is taken ``mod 1":
\bsplit{nearResonance}
	\xi_k &= \xi_0 +  \hat{\Omega}^* + \eps^{p} \lambda \zeta_k 
	             +\cO(\eps, \eps^{2p}) \;,\\
	\eta_k  &= \eta_0 + \eps^p \alpha \zeta_k + \eps ^{2p}(\beta \zeta_k^2 -\delta) 
				+\cO(\eps,\eps^{3p}) \;,\\
	\zeta_k&= \zeta_0 + \eps^{1-p} \hat g(\xi_0,\eta_0) + \cO(\eps)  \;,
\end{split} \end{equation}
where
\[
	\hat g(\xi,\eta) = \sum_{j=0}^{k-1} g(\xi+\tfrac{j}{k}\hat\Omega^*,\eta+\tfrac{j}{k}n) \;.
\]
This near-resonance map is partitioned into {\em fast}, $\xi$,  and {\em slow}, $(\eta,\zeta)$, variables. 

Note that \Eq{nearResonance} is essentially the same as \Eq{stdTangency}, though with a more general force $g$. That is, if one neglects the higher-order terms in \Eq{nearResonance} and then formally undoes the scaling, by setting $(x,y,z) = (\xi, \eta, \eps^{p}\zeta)$, the result is a $d$-dimensional extension of \Eq{stdTangency} (with an additional parameter $\lambda$ that can be also removed by scaling when it is nonzero). This reinforces calling \Eq{stdTangency} the ``standard volume-preserving" map.

Since $\Omega^*$ satisfies precisely one resonance condition and the columns of $\hat{M}^T$ are independent of $\hat m$, whenever $l \in \bZ^{d-1}$ is nonzero $l \cdot \hat{\Omega}^* = kl\cdot \hat M \Omega^* = k (\hat M^T l) \cdot \Omega^* \notin \bZ$. In particular none of the components of this transformed frequency can be rational: the angles $\xi$ are ``rapidly varying." If, in addition, $\hat \Omega^*$ is Diophantine, \Thm{Averaging} implies that,
on the time scale $t = \cO(\eps^{p-1}) \gg 1$, the fast $\xi$ angles can be averaged away, reducing the system \Eq{nearResonance} to the two-dimensional map
\bsplit{twoDAveraged}
	\eta_k  &= \eta_0 + \eps^p \alpha \zeta_k + \eps ^{2p}(\beta \zeta_k^2 -\delta) 
				+\cO(\eps,\eps^{3p}) \;,\\
	\zeta_k&= \zeta_0 + \eps^{1-p} \bar g(\eta_0) + \cO(\eps)  \;,
\end{split}\end{equation}
where
\[
	\bar g(\eta) =  \int_{\bT^{d-1}} \hat g(\xi,\eta) d\xi
	                    =  \sum_{j=0}^{k-1} \int_{\bT^{d-1}} 
	                       g(\xi+ \tfrac{j}{k} \hat \Omega^*,\eta+\tfrac{j}{k}n) d\xi
				  = k\sum_{l\in \bZ} \tilde g_{0,kl} e^{2\pi i kl\eta} \;, 
\]
contains only the Fourier coefficients of $g$ that are multiples of $(0,k)$. Thus
\[
	\bar g\left(\eta+ \tfrac{1}{k}\right) = \bar g(\eta) \;.
\]

Note that \Eq{twoDAveraged} is area preserving. In \Sec{Averaging} we will show that
this property holds more generally, and that the degree of approximation between the 
averaged and the original system can be increased with additional coordinate 
transformations.

To lowest order, the averaged map can be related to a Hamiltonian flow. 
For this to be possible we assume that $\tilde{g}_0 = 0$, 
i.e., that $\bar g$ has zero average. Even though the derivation of \Eq{twoDAveraged} is valid when this is not true, the resulting area-preserving map would then have nonvanishing net flux and consequently, no invariant tori; it also would not be related to the flow of a Hamiltonian. For the rest of this section we assume that the average of $Z$ is zero, as we also did in \Sec{Example}.

There are two distinguished limits corresponding to transversality or tangency of the frequency curve with the resonant plane; these can be treated by selecting the exponent $p$. If $\alpha$, \Eq{alphabeta}, is nonzero, the frequency curve $\Omega$ crosses the resonance transversely (since the vector $m$ is perpendicular to the resonant plane, and $D\Omega$ is tangent to the frequency curve). In this case we set $p = \tfrac12$ and the averaged map \Eq{twoDAveraged} is approximately the time $\sqrt{\eps}$ flow map of the Hamiltonian
\[
	H(\eta,\zeta) = \frac{\alpha}{2} \zeta^2 + V(\eta) \;,
\]
with $DV(\eta) = -\bar{g}(\eta)$. This is similar to the standard pendulum \Eq{Pendulum} and gives a resonance of size $\sqrt{\eps}$ about $z^*$ in the original coordinates, as discussed in \Sec{Example}.

On the other hand, if $\alpha = 0$, the frequency curve is tangent to the resonance plane. If $\beta$, \Eq{alphabeta} is nonzero, then its curvature at the tangency is nonzero. In this case we set $p = \tfrac13$ and \Eq{twoDAveraged} limits on the time $\eps^{\frac23}$ flow map of the Hamiltonian
\[
	H_R(\eta,\zeta) = \frac{\beta}{3} \zeta^3 -\delta \zeta + V(\eta) \;.
\]
This is similar to the standard reconnecting Hamiltonian \Eq{HReconnect}.

We have already studied in detail the two cases $\alpha \neq 0$ and $\alpha = 0$, $\beta \neq 0$ and their corresponding averaged Hamiltonians in \Sec{Example}. Consequently, these examples describe the typical structure of the dynamics near a rank-one resonance for the cases of transverse and quadratically tangent resonances.
Of course, in \Sec{Example} we only considered the dynamics with a specific, simple choice of $g$. What is surprising is that the numerical comparisons in \Sec{Example} indicate that the averaging theory (which is only valid for time of order $\eps^{p-1}$) nevertheless gives a qualitative understanding (to the level of resonance structure, widths and the onset of chaos through overlap) of the long time dynamics. There will, of course, be many features that this simple theory misses.

\section{Averaging for Volume-Preserving Maps with Nearly Constant Frequency}\label{sec:Averaging}
When a dynamical system can be separated into slow and fast components, averaging can often be used to describe the slow dynamics on a long time scale \cite{Lochak88, Sanders07}. In this section we prove an averaging theorem relevant to the system \Eq{nearResonance} for which the fast dynamics, the $\xi$ variables, corresponds to translations on a torus with nearly constant frequency.  
In fact, the theory applies to more general maps with any number of fast and slow variables. 
To reinforce this, we denote the fast angles by $\theta$, and the slow variables ($(\eta,\zeta)$ in the previous section) 
by $J$---to indicate their similarity to ``actions". 
The slow-fast system (in particular also the map \Eq{nearResonance}) can then be written 
\bsplit{weakCoupled}
	\theta' &= \theta + \omega_0 + \kappa F^{(1)}(\theta,J;\kappa) \;,\\
	J'      &= J + \kappa G^{(1)}(\theta,J;\kappa) \;,
\end{split}\eeq	
on  $\bT^l \times M$ where $M \subset \bR^j$ for some $j$. Here $\kappa$ represents the small parameter $\eps^{p}$ of \Eq{nearResonance}.

We will show, in addition, that if the original system is volume preserving, 
then the coordinate transformations that achieve the (higher-order) averaging can be 
chosen to be volume preserving as well. 
The implication is that the truncated, averaged system will have a symmetry (i.e., 
translation along the fast angles) and thus the decoupled, slow system will again be 
volume preserving. In our particular example this shows that the map \Eq{twoDAveraged}, 
in appropriately transformed slow variables $(\eta, \zeta)$, is area preserving.

The goal of the averaging is to obtain a description of the slow dynamics on the time scale $\cO(\kappa^{-1})$. One hopes that if  $\kappa \ll 1$ and $\omega_0$ is sufficiently incommensurate, then the slow dynamics is approximated by 
\beq{firstOrderAveraged}
	\bar J' = \bar J + \kappa \bar G^{(1)}(\bar J) \;,
\eeq
where
\beq{barDef}
	\bar G(J) \equiv \int_{\bT^l} G(\theta,J;0) d\theta \;.
\eeq
This first order averaged equation should give dynamics close to that of the full system in the sense that for any orbit $(\theta_t,J_t)$ of \Eq{weakCoupled}, then provided that $\bar J_0 = J_0$ we should have
\[
	|\bar J_t - J_t| = \cO(\kappa)
\]
for $0 < t < \frac{T}{\kappa}$ for any fixed $T > 0$.

In most cases averaging theory is formulated for ODEs, though there are some results specifically for maps \cite{Kifer03, Dumas04, Brannstrom09}. The analyses of Kifer and Br\"annstr\"om apply to systems in which the slow and fast dynamics are more strongly coupled than \Eq{weakCoupled} and for which the slow dynamics need not correspond to toral translations. In \cite{Brannstrom09} the angle dependence of $G$ is assumed to occur only at $\cO(\kappa^2)$. These theorems apply only when the constant vector $\omega_0$ is replaced by a function $\Omega(J)$ that satisfies a nondegeneracy, or twist condition (Kolmogorov nondegeneracy).  In either case, they must deal with the difficulty that the ergodicity of the fast angle dynamics can depend, at lowest order, on the slow variables; consequently in order to apply these theorems, a nontrivial, ``good-ergodization" property must be verified. The results are  weak in the sense that the averaged and true, slow dynamics remain close only in the sense of measure.
The reason we can achieve better results is that we are treating the simple case where the frequency of the unperturbed system is constant and Diophantine. Because of this we are able to achieve arbitrary high order accuracy for every initial condition.

A standard implementation of averaging begins with the construction of a coordinate change to push the coupling to the fast angles off to higher order \cite{Sanders07}. 
Being ultimately only interested in the slow dynamics, we could choose not to transform 
the angle dynamics. However, we will see that the truncated, slow map
is volume-preserving if the fast angle variables are transformed as well.
When $\omega_0$ is Diophantine, \Eq{Diophantine}, and the map is analytic, this transformation can be done recursively to any order. We start by assuming that the normalizing transformation has been done up to $\cO(\kappa^{n-1})$, but the result also applies to \Eq{weakCoupled} directly by setting $n = 1$.

\begin{lem}[Decoupling Transformation]\label{lem:Transformation}
Let $f$ be the real analytic map on $\bT^l \times M$
\begin{align}
	\theta' &= \theta + \omega_0 + 
	                     \sum_{j=1}^{n-1}\kappa^j \bar F^{(j)}(J) + \kappa^n F^{(n)}(\theta,J;\kappa) \;,\label{eq:thetaMap}\\
	J'      &= J + \sum_{j=1}^{n-1}\kappa^j \bar G^{(j)}(J)  +\kappa^n G^{(n)}(\theta,J;\kappa)\;, \label{eq:JMap}
\end{align}
such that $\omega_0 \in \cD(c,\mu)$ for some $c,\mu >0$. Then there is a $\kappa_0 >0$ such that for all $|\kappa| < \kappa_0$ there exists a diffeomorphism $h^{(n)}(\theta,J) = (\psi,I)$ (inversely) defined by  
\beq{transform}
	(\theta,J)  = (\psi + \kappa^n T^{(n)}(\psi,I), I + \kappa^n S^{(n)}(\psi,I) ) \;,
\eeq
that conjugates the dynamics to the analytic map
\bsplit{orderNp1}
	\psi' &= \psi + \omega_0 + 
	           \sum_{j=1}^{n}\kappa^j \bar F^{(j)}(J) 
	           + \kappa^{n+1} F^{(n+1)}(\psi,I;\kappa) \;, \\
	I'  & = I + \sum_{j=1}^{n}\kappa^j \bar G^{(j)}(I)  
	           +\kappa^{n+1} G^{(n+1)}(\psi,I;\kappa)\;, 
\end{split}\eeq
where $\bar G^{n}$ and $\bar F^{(n)}$ are the averages of $G^{(n)}$ and $F^{(n)}$, respectively, as defined by \Eq{barDef}.
\end{lem}

\begin{proof}
The elimination of $\theta$-dependent terms to $\cO(\kappa^n)$ proceeds exactly the same 
for the $\theta$ and $J$ equations. We present the argument for \Eq{JMap}---replacing $(G,S)$ by $(F,T)$ would give the argument for \Eq{thetaMap} (the affine term, $\omega_0$, simply carries through).

To eliminate $\theta$-dependent terms substitute \Eq{transform} into \Eq{JMap} 
and expand to $n^{th}$ order to obtain the \emph{homological} equation
\beq{homological}
    S^{(n)}(\psi+\omega_0,I) - S^{(n)}(\psi,I) = G^{(n)}(\psi,I;0) - \bar G^{(n)}(I) \;.
\eeq

Formally, this can be solved by Fourier transformation, setting
\[
	S^{(n)}(\psi,I) = \sum_{k \in \bZ^l} \tilde S^{(n)}_k(I) e^{2\pi i k \cdot \psi} \;,\quad
	G^{(n)}(\psi,I;0) = \sum_{k\in\bZ^l} \tilde G^{(n)}_k(I) e^{2 \pi ik\cdot \psi} \;,
\]
and noting that $\tilde G^{(n)}_0(I) = \bar G^{(n)}(I)$ 
to obtain
\beq{fourier}
	\tilde S^{(n)}_k(I) = \frac{ \tilde G^{(n)}_k(I)}{ e^{2 \pi i k \cdot \omega_0} -1} \;, \; k \neq 0\;.
\eeq
The function $\tilde S^{(n)}_0(I) = \bar S^{(n)}(I)$ can be freely chosen, for example, to be $0$.

Since $\omega_0$ is incommensurate, the denominator of \Eq{fourier} never vanishes: each $\tilde S_k(I)$ is an analytic function of $I$. Moreover, when $\omega_0$ is Diophantine  and  $G$ is analytic in $\theta$ then the formal Fourier series for $S$ converges to an analytic function, see for example \cite{Moser66}.

The new map for $I$ is 
\bsplit{yMap}
	I' &= J' - \kappa^n S^{(n)}(\psi',I') \\
       & = I +\sum_{j=1}^{n-1}\kappa^j \bar G^{(j)}(J) 
           +\kappa^n[ G^{(n)}(\theta,J;\kappa) + S^{(n)}(\psi,I)- S^{(n)}(\psi',I') ] \;.
\end{split}\eeq 
Comparing with \Eq{orderNp1}, gives an implicit definition for $G^{(n+1)}$:
\bsplit{Gnplus1}
	G^{(n+1)}(\psi, I;\kappa) =  \frac{1}{\kappa}
	    & \left[\sum_{j=1}^{n-1} \kappa^{j-n} \left\{ 
	        \bar G^{(j)}(J)-  \bar G^{(j)}(I)\right\} +  
	    G^{(n)}(\theta, J;\kappa) - \bar G^{(n)}(I) +\right.\\
	 &\left. S^{(n)}(\psi,I)
	   	    -S^{(n)}\left(\psi', I'\right)\right] \;,
\end{split}\eeq
where $J$ and $\theta$ are just short-hand notation for  the right hand side of 
\Eq{transform}. A similar equation is obtained for $F^{(n+1)}$ and these equations 
are implicit and coupled because $(I', \psi')$ is a function of $(G^{(n+1)}, F^{(n+1)})$. 
The homological equation \Eq{homological} implies that $G^{(n+1)}$ is well-defined as $\kappa \to 0$,
because it forces the $O(\kappa^0)$ term inside the square bracket in \Eq{Gnplus1} to vanish.
The leading order term at $\kappa = 0$ thus is
\begin{align*}
	G^{(n+1)}(\theta,I;0) &= D_I \bar G^{(1)}(I)S^{(n)}(\theta,I)  + D_\kappa G^{(n)}(\theta,I;0) \\
	      &- D_\theta S^{(n)}(\theta + \omega_0,I)\bar F^{(1)}(I)
	      - D_IS^{(n)}(\theta+\omega_0, I)\bar G^{(1)}(I) \;.
\end{align*}
Moreover, since the Jacobian of the right hand side of \Eq{Gnplus1} and the analogous equation for $F^{(n+1)}$ 
with respect to $(G^{(n+1)}$, $F^{(n+1)})$ vanishes at $\kappa = 0$, the implicit function theorem implies that there is a $\kappa_0>0$ such that for all $|\kappa| < \kappa_0$, the functions $G^{(n+1)}$,  $F^{(n+1)}$ exist and are analytic.

Finally, since the transformation \Eq{transform} is a near-identity transformation, $\kappa_0$ can be selected so that it is a diffeomorphism. Thus the dynamics of $(\psi,I)$ and $(\theta,J)$ are diffeomorphically conjugate.

\end{proof}




We now show that \Eq{transform} can be replaced to by a volume-preserving transformation 
when $f$ is itself volume preserving. In order to do this we interpret $(T^{(n)},S^{(n)})$ 
as a vector field $V = (T^{(n)} \partial_\psi, S^{(n)} \partial_I)$. This vector field 
generates a transformation $e^{-V} (\psi,I)$ that is volume preserving whenever $V$ is 
divergence free, $\nabla \cdot V = 0$. To show that this is possible, we first prove two 
simple lemmas. The first exploits the assumption that the Jacobian of 
\Eq{weakCoupled} has the form $Df = I + \kappa B_1 + \cO(\kappa^2)$. 
If $f$ is volume preserving, then $1 = |Df| = 1 + \kappa \tr B_1 + \cO(\kappa^2)$, 
which implies $\tr B_1 = 0$. Note that higher order terms do not necessarily have 
zero trace, however. Put differently, when $\tr B = 0$, then $| \exp B | = 1$, 
but only the linear term in $\exp B = 1 + B + \frac12 B^2 + \dots$ necessarily has 
vanishing trace. More generally:

\begin{lem} \label{lem:det1}
Consider a convergent sequence of nonsingular, square matrices $M_n = \sum_{k=0}^n \kappa^k B_k$ %
such that  the determinant $| M_\infty| = 1$.
Then for any finite $n$ we have $\tr B_n = C_n(B_{n-1}, \dots, B_1)$ where $C_n$ is some 
smooth function of the matrices $B_k$ with indices $k < n$.
\end{lem}

\begin{proof}
Firstly, writing $M_\infty = M_{n-1} + \cO(\kappa^{n})$ and taking the 
determinant of both sides gives
$1 = |M_{n-1}| |I + \cO(\kappa^{n})| = |M_{n-1}| + \cO(\kappa^{n}) $, which implies  
\[
    |M_{n-1}| = 1 + \cO(\kappa^{n}) = 1 - \kappa^{n} C_{n}( B_{n-1}, B_{n-2}, \dots, B_1) + \cO(\kappa^{n+1}) \;.
\]
Secondly, observing that  $M_n = M_{n-1} + \kappa^n B_n$ and taking determinants of this 
identity gives $|M_n| = |M_{n-1}| | I + \kappa^n M_{n-1}^{-1} B_n |$.  
Using the previous estimates for the determinants of $M_n$ and $M_{n-1}$ then implies
\[
    1 + \cO(\kappa^{n+1}) = (1 - \kappa^n C_n + \cO(\kappa^{n+1}))
      (1 + \kappa^n \tr B_n + \cO(\kappa^{n+1})) \;,
\]
where in the last term we have used that $M_{n-1} = I + \cO(\kappa)$ so that 
$M_{n-1}^{-1} = I + \cO(\kappa)$  and hence $\tr ( M_{n-1}^{-1} B_n )  = 
\tr B_n + \cO(\kappa)$. Finally, expanding the right hand side and equating 
terms of order $\cO(\kappa^n)$ in this equation gives $\tr B_n = C_n$.
\end{proof}

Next we establish a connection between the divergence of a periodic vector field 
and the divergence of its average.

\begin{lem} \label{lem:avgG}
Let $V$ be a vector field on $T^l \times M$ in coordinates $(\theta, J)$ 
and denote by a bar the average over all angles $\theta$. 
If $\nabla \cdot V = C(J)$ where $C$ is a function of $J$ only, 
then $\nabla \cdot V = \nabla \cdot \bar V$.
\end{lem}
\begin{proof}
First we show that in general $\overline{ \nabla \cdot V} = \nabla \cdot \bar V$.
Write $V = (F, G)$ where $F : T^l \times M \to T^l$ and $G : T^l \times M \to M$ so that $\nabla \cdot V = \tr D_\theta F + \tr D_J G$. Since $F$ is periodic in 
$\theta$, the average of any $\theta$-derivative vanishes. Hence 
$\overline{ \nabla \cdot V} = \tr \overline{ D_J G} = 
\tr D_J \bar G = \nabla \cdot \bar V$ since the operations of averaging 
over $\theta$ and differentiation with respect to $J$ commute. 
By assumption we have $\nabla \cdot V = C(J)$ and averaging this identity gives 
$\overline{ \nabla \cdot V}  = \overline{C(J)} = C(J)$.
Finally, replacing $\overline{ \nabla \cdot V}$ by $\nabla \cdot \bar V$ and eliminating $C(J)$ gives the result.
\end{proof}

With these two lemmas we can now prove:

\begin{lem}[Volume-Preserving Decoupling Transformation]\label{lem:VPDecoupling}
If the map $f$ is volume-preserving then the decoupling transformation 
\Eq{transform} can be extended to a volume-preserving map.
\end{lem}
\begin{proof}
The volume-preserving extension of the near-identity transformation \Eq{transform} is constructed by interpreting $V = (T^{(n)} \partial_\psi, S^{(n)} \partial_I)$ as a vector field and using its exponential as the transformation operator. What we need to show is that if the map $f$ is volume-preserving then $V$ can be chosen to be divergence free,
so that the resulting exponential is volume-preserving. 

The first observation is that the homological operator \Eq{homological} considered as acting on vector fields, preserves the subspace of divergence free vector fields and its complement. Introducing the unperturbed map $h_{\omega_0}(\theta, J) = (\theta + \omega_0, J)$ the homological operator can be written as
\[
   L_{\omega_0}V = (D h_{\omega_0})^{-1} V \circ h - V 
                     = (h^*_{\omega_0} - id) V \;.
\]
Now if $\nabla \cdot V = 0$, then this also true for the 
image $L_{\omega_0}(V)$ because $h_{\omega_0}$ is volume-preserving. 
Moreover, since $L_{\omega_0}$ is semi-simple, the complement of any 
invariant subspace is preserved as well.

To complete the proof, we need to show that the right hand side 
of the homological equation,
$(F^{(n)}(\psi,I;0) - \bar F^{(n)}, G^{(n)}(\psi,I;0) - \bar G^{(n)})$, is a 
divergence free vector field, so that the generator of the transformation 
$(T^{(n)},S^{(n)})$ can be chosen to be divergence free as well. 

To apply \Lem{det1} we identify the Jacobian of \Eq{thetaMap} and 
\Eq{JMap} with $M_n$, i.e., $B_j = D(\bar 
F^{(j)}, \bar G^{(j)})$ for $j = 1, \dots, n-1$ and set 
$B_n = D(F^{(n)}(\psi, I;0), G^{(n)}(\psi, I;0)$. 
The result is that $\tr B_n = \nabla \cdot ( F^{(n)}(\psi, I;0),  G^{(n)}(\psi, I, 
0)) = C_n(B_{n-1}, \dots, B_1) = C_n(I)$ since all lower order terms have 
already been averaged. This sets the stage for \Lem{avgG}, which implies 
that $ \nabla \cdot ( F^{(n)}(\psi, I;0),  G^{(n)}(\psi, I;0)) = 
\nabla \cdot (  \bar F^{(n)}, \bar G^{(n)})$, so that the divergence of the 
difference $(  F^{(n)} - \bar F^{(n)} ,  G^{(n)} - \bar G^{(n)})$ vanishes. 
Therefore, the right hand side of the homological equation is divergence free, 
and hence the generator $(T, S)$ can also be chosen to be divergence free at each stage. 

Finally, if we replace the transformation \Eq{transform} by $(e^{T^{(n)}\partial_\psi} \psi , e^{S^{(n)}\partial_I} I)$, we obtain the result.
\end{proof}

Now we return to the general (not necessarily volume-preserving) case.
We will show that, roughly speaking, the projection $\pi (\psi,I) = I$ of the slow dynamics of \Eq{weakCoupled} onto $M$ is conjugate,
to $\cO(\kappa^{n})$, to the averaged map
\beq{averaged}
	\bar I' = \bar I + \sum_{j=1}^{n} \kappa^j \bar G^{(j)}(\bar I)
\eeq
 for times of order $\kappa^{-1}$ for any $n \ge 1$.
 More precisely:

\begin{thm}[$n^{th}$ Order Averaging]\label{thm:Averaging}
Suppose that $f$ is an analytic map of the form \Eq{weakCoupled} and $\omega_0$ is Diophantine. 
Let $h = h^{(n)} \circ h^{(n-1)}\circ \ldots h^{(1)}$ be the composition of the sequence of transformations,
\[
	(\psi, I ) = h^{(j)}(\theta, J) \;,
\]
given by the inverse of \Eq{transform}, which exists for $|\kappa| < \kappa_0$ for 
some $\kappa_0 > 0$ by \Lem{Transformation}.
If given a $T>0$, the orbit  $\bar I_t$ of  \Eq{averaged} is in some compact subset $K \subset M$ for $0 < t < \frac{T}{\kappa}$, then
\beq{orderNBound}
	| \pi h(\theta_t,J_t) - \bar I_t| = \cO(\kappa^{n}) \;,
\eeq
where $(\theta_t, J_t)$ is {\em any} orbit of $f$ with initial conditions satisfing $\bar I_0 = \pi h(\theta_0,J_0)$.
\end{thm}

\begin{proof}
By \Lem{Transformation}, $h$ conjugates \Eq{weakCoupled} to \Eq{orderNp1}. Thus $\pi h(\theta_t, J_t) = I_t$, an orbit of the second component of \Eq{orderNp1}.
Since $G^{(n+1)}$ is analytic on $\bT^l \times K$ and $K$ is compact, it is bounded, say $|G^{(n+1)}| \le C$. 
Similarly, on $K$ the function 
$\sum_{j=1}^{n}\kappa^j\bar G^{(j)}(I)$ is Lipschitz, say with constant $\kappa L$. Consequently, 
\[
	|I_t - \bar I_t| \le  \kappa |I_{t-1}-\bar I_{t-1}| + 2C\kappa^{n+1} \;,
\]
and  the discrete Gr\"onwall lemma, e.g.~\cite{Argarwal00}, implies that
\[
	| I_t - \bar I_t|  \le 2C\kappa^{n+1} t (1+\kappa L)^{t-1} \le 
	  \kappa^n (2CT  e^{LT})
\]
for all $0 < t < \frac{T}{\kappa}$.
\end{proof}

\begin{rem}
It may seem difficult to find an appropriate set $K$ that contains the orbits of
interest over time $T/\kappa$ in the limit of vanishing $\kappa$.
For this it is useful to notice that for sufficiently small $\kappa$ the orbits of $\bar I_t$ are well approximated by the flow of the divergence free vector field 
$\bar G^{(1)}$ for time $T$. If this time-$T$ flow is bounded,
then the region of $M$ covered by the flow is a good candidate for the set $K$.
\end{rem}

In \Thm{Averaging}, the map was normalized up to $\cO(\kappa^{n})$ with remaining 
correction terms $(F^{(n+1)}, G^{(n+1)})$ of order $\cO(\kappa^{n+1})$.
In practice it is easy to compute the averages $(\bar F^{(n+1)}, \bar G^{(n+1)})$, 
which gives the next order term of the averaged map \Eq{averaged} whose orbits 
we denote by $\bar I^{(n+1)}_t$. It turns out that it is not necessary to compute
the next order transformation $h^{(n+1)}$ to get an error of order $\kappa^{n+1}$.

\begin{cor} \label{cor:Averaging}
Denote by $\bar I^{(n+1)}_t$ be an orbit of \Eq{averaged} with $n$ replaced by $n+1$.
With the other assumptions as in \Thm{Averaging} we then have
\[
   | \pi h(\theta_t, J_t ) - \bar I^{(n+1)}_t | = \cO( \kappa^{n+1}) \;.
\]
\end{cor}

\begin{proof}
If we apply \Thm{Averaging} at order $n+1$ the statement is
\[
   | \pi (h^{(n+1)} \circ h)(\theta_t, J_t ) - \bar I^{(n+1)} | = \cO( \kappa^{n+1}) \;.
\]
Observing that $h^{(n+1)} = id + \cO(\kappa^{n+1})$ and expanding the right hand side gives the result.
\end{proof}

Combining $n$th order averaging, \Thm{Averaging}, and volume-preserving decoupling, \Lem{VPDecoupling}, we find:

\begin{cor}[Volume-Preserving Slow Map]\label{cor:VPSlow}
If the map \Eq{weakCoupled} is volume-preserving, then the transformation of \Lem{Transformation} can be done so that the truncated slow map \Eq{averaged} is volume-preserving on $M$ up to order $n$.
\end{cor}

\begin{proof}
After decoupling and truncating at order $n$, the transformed map \Eq{orderNp1} is independent of $\psi$  (except for its identity part). 
Consequently, the Jacobian of the truncated map is upper block-diagonal, and the upper left block is the identity. This results from the translation symmetry of \Eq{averaged} along $\psi$, and its semidirect product form.
Thus the determinant is equal to the determinant of the lower right block, which is the 
Jacobian of the slow map \Eq{averaged} alone. 
As in \Lem{VPDecoupling}, the decoupling transformation $h$ can be chosen to be volume preserving, and so the determinant of the Jacobian of the untruncated map is equal to 1. When terms of order higher than $n$ are neglected the determinant is $1 + \cO(\kappa^{n+1})$.
\end{proof}

\begin{rem} One may choose to leave the fast dynamics untransformed, setting $\theta = \psi$ in each decoupling step. This would still decouple the slow dynamics from the fast dynamics, and \Thm{Averaging} still holds. However, the reduced map would not in general be volume preserving.
\end{rem}

\begin{figure}
\centering{\includegraphics[width=16cm]{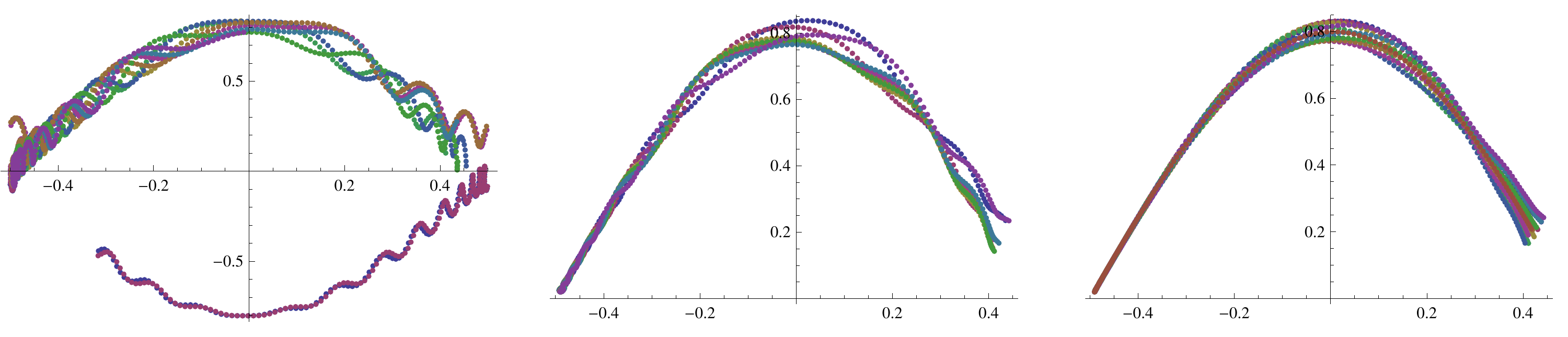}}
\centering{\includegraphics[width=16cm]{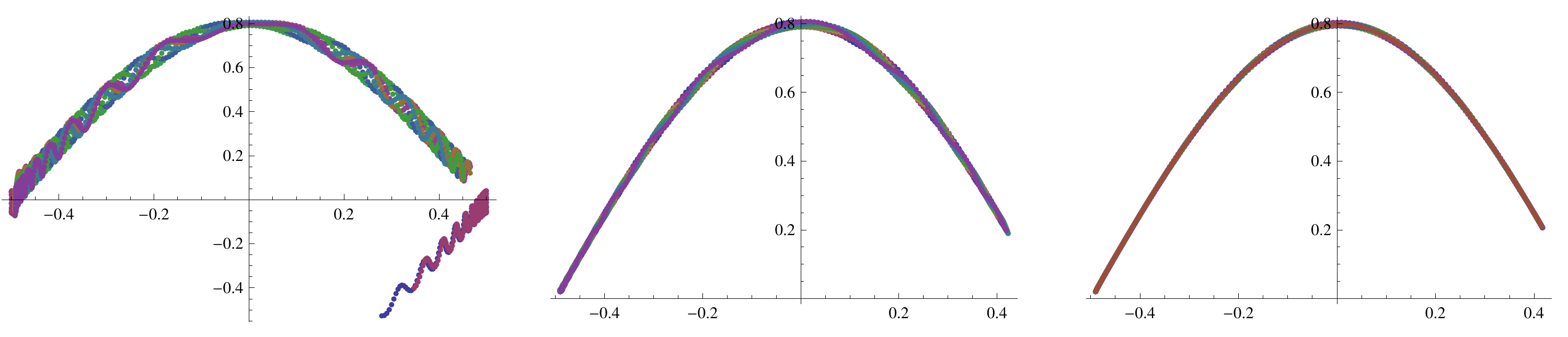}}
\caption{Projections of ten orbits of \Eq{stdTangency} with initial conditions $x=J_1 = -0.49$ and $\zeta =J_2 = 0.02$ and covering $y = \theta$ in equidistant steps.
Parameters are chosen as in \Fig{TransverseResonances} except $\delta = 0.7$.
The panes in a row show $\pi h(\theta_t, J_t)$ for $h = id$, $h = h^{(1)}$, $h = h^{(2)} \circ h^{(1)}$ from left to right, respectively.
For the top row
$\eps = \kappa^2 = 0.0002$, $t = 0, \dots, 177$,
while for the bottom row 
$\eps = \kappa^2 = 0.0001$, $t = 0, \dots, 250$.
The averaged orbit, of \Eq{stdMap}, is essentially identical to the lower right plot.
}
 \label{fig:averaging}
\end{figure}

We now give some details of this averaging procedure applied to 
the example \Eq{stdTangency} with perturbation \Eq{Froeshle} 
near the $(1,0,n)$ resonance, where $z^* = n -\gamma$.
In this case the function $g$ is a trigonometric polynomial. After blowing-up near the $(1,0,n)$ resonance, the map \Eq{StandardZoomed} has fast variable $\theta = \xi$ with $\omega_0 = \beta(n-\gamma)^2 - \delta$, slow variables $J = (\eta,\zeta)$, and small parameter $\kappa = \sqrt{\eps}$. The fast dependence occurs in every component of the map; however, in the $\xi$ and the $\eta$ equations, it only appears at quadratic order in $\kappa$. As a consequence, the first order decoupling transformation \Eq{transform} only changes $\zeta$.

Since $g$ is already given as a Fourier series it is easy to compute the transformation:
\beq{S1trafo}
   [T^{(1)},S^{(1)}] = \left[0,0,\, \frac{1}{2 \sin( \pi \omega_0)} ( b \cos ( 2 \pi \xi - \pi \omega_0) - c \cos( 2 \pi (\xi-\eta) - \pi \omega_0)\right] \;.
\eeq
Note the occurrence of the ``small-denominator" $\sin(\pi \omega_0)$ that arises from \Eq{fourier}. Since the $\cO(\kappa)$ terms in $\zeta'$ are independent of $\zeta$,
$S^{(1)}$ is independent of $\zeta$,
and the resulting transformation is exactly volume-preserving. In the new variables the dependence on the fast angle $\xi$ appears at $\cO(\kappa^2)$ in every equation.
Averaging these equations over $\xi$ gives the standard map \Eq{stdMap}.

The next order transformation depends on all variables, and has all components non-zero:
\[
   \begin{pmatrix} T^{(2)} \\ S^{(2)} \end{pmatrix} = \frac{1}{4 \sin^2 \pi \omega_0}
    \begin{pmatrix}
          2 \beta(n-\gamma) \big(b \sin 2 \pi \xi + c \sin 2\pi (\eta - \xi)\big) \\
          b \sin 2\pi \xi + c \sin 2\pi(\eta-\xi)  \\
          2\pi \zeta(2b \beta(\gamma-n) \cos 2\pi \xi 
              - c(1 + 2 \beta(\gamma-n))\cos2\pi(\eta-\xi) )
      \end{pmatrix} \;.
\]
The corresponding transformation removes the $\xi$ dependence up to $\cO(\kappa^2)$. 
Averaging the transformed equations at $\cO(\kappa^3)$ gives a correction to the 
standard map that has the sole effect of changing the coefficient of 
$\sin 2\pi x$ from $-\kappa a$ to 
\[
     -\kappa \tilde a = -\kappa a - \kappa^3 \frac{bc\pi}{4 \sin^2 \pi \omega_0 }   \;.
\]
At the next order, additional terms of order $\kappa^4$ in the truncated decoupled map are introduced and the final averaged map becomes
\begin{align*}
     I_1' &= I_1 + \kappa I_2' + \kappa^4 \chi(I_1) \;,\\
     I_2' &=  I_2 - \kappa \tilde a \sin 2\pi I_1 - \kappa^4  I_2 D \chi(I_1) \;,
\end{align*}
where $\chi(I_1) \equiv \gamma_1 + \alpha_1 \sin2\pi I_1 + \beta_1 \cos 2\pi I_1$.

The orbits of the full map in the transformed coordinates are shown in \Fig{averaging}, showing that as the order of the transformation is increased, or the size of the perturbation is decreased, the fluctuations due to the fast dynamics become smaller. Log-log plots of error vs.\ $\kappa$ clearly verify that the scaling of the deviation from the averaged orbits has exponents $2$, $3$, $4$ as promised by \Cor{Averaging}.

\section{Double Resonance}\label{sec:RankTwo}

For the one-action map \Eq{integrable} in three dimensions, the image of the frequency map
is a two-dimensional frequency vector $\omega$. If its resonance has rank two, then $\omega$ must be rational, say,
\[
	\omega = \left( \frac{l_1}{d_1}, \frac{l_2}{d_2} \right) \;.
\]
Note that orbits of the integrable torus map \Eq{torusMap} with this rational frequency are periodic with period $\lcm(d_1,d_2)$. In this case, the near-integrable map \Eq{Perturbed} has two slow angles, and as we will see, the map cannot be reduced in dimension by averaging.

To obtain the normal form, we first address the problem of finding an appropriate basis for the module $\cL(\omega)$. Note that, without loss of generality, we can assume that $gcd(l_i, d_i) = 1$; however, the denominators $d_1$ and $d_2$ need not be coprime.
When $gcd(d_1, d_2) = 1$ the only way to achieve $m \cdot \omega \in \bZ$ is to cancel the 
denominators $d_i$, implying that the resonance module has a basis $m_1 = (d_1, 0)$ and $m_2 = (0, d_2)$. By contrast, when
\[
	g \equiv gcd(d_1, d_2) \neq 1
\]
the problem of finding a basis for $\cL(\omega)$ is more interesting. 
One basis vector is easy, either $(d_1, 0)$ or $(0, d_2)$ suffices. 
However, together these two vectors do not form a basis, since there is additional cancellation. If for concreteness we choose $m_1 = (d_1,0)$, then upon
writing $d_i = g \hat d_i$, the equation to solve for the second vector $m_2 = (p,q)$ becomes
\[
   g m_2 \cdot \omega = \frac{l_1 }{\hat d_1} p + \frac{l_2 }{\hat d_2} q = g n, 
   		\quad p,q, n \in \bZ \;.
\]
This implies that $p = \hat d_1 i$ and $q = \hat d_2 j$, so we arrive at 
\[
   l_1 i + l_2 j = g n, \quad i, j \in \bZ \;,
\]
where by construction $gcd( l_i, g) = 1$.
To make $m_2$ independent of $m_1$, we must choose $j \ne 0$.
We can always find a solution with $j=1$, since the Diophantine equation 
\[
   g n - l_1 i  = l_2 \;
\]
is always solvable for integers $i$ and $n$ when $gcd( g, l_1) = 1$ \cite{HardyWright79}. Thus a second basis vector is $m_2 = (i\hat d_1, \hat d_2)$.
Note that the determinant of the matrix of basis vectors is $d_1 \hat d_2 = g \hat d_1 \hat d_2 = \lcm(d_1, d_2)$, which is the period of orbits with rotation vector $\omega$.

\begin{example}
	The module for $\omega = (\frac13, \frac 12)$ has $g = 1$, so we can take
	$m_1 = (3,0)$ and $m_2 = (0,2)$ and then
	\[
		\cL = \{(3i,2j), (i,j)\in \bZ^2 \} \;.
	\]
	Of course any unimodular transformation of $m_1$ and $m_2$ also gives a basis;
	however, the selected basis is minimal in the sense that no other basis 
	has smaller $1$-norm.
\end{example}

\begin{example}
	The module for $\omega = (\frac12, \frac14)$ has period
	$4$ and $g = 2$.  In this case the choice $m_1 = (2,0)$ and $m_2 = (0,4)$
	misses some lattice points, for example $(1,2)$. 
	The process outlined above gives $m_2 = (1,2)$ and the same $m_1$,
	which is the minimal basis in the $1$-norm.
\end{example}

\begin{example}
	If $\omega = (\frac{1}{12}, \frac19)$ then the period is $36$ and $g = 3$. If
	 $m_1 = (12,0)$, we must solve the equation $i+j = 3n$.
	 For $j=1$ a solution is $i=2$, so we have a second vector $m_2 = 
	 (8,3)$. The minimal basis is $ (4, -3)$, $(8,3)$. 
\end{example}

Once a basis for the module is known, the analysis of \App{Integrable} can be used to choose a set of angles on the torus that have diagonal dynamics. In particular, we can diagonalize the angle dynamics using the transformation \Eq{diagonalization},
$M = PKQ$ where $P$ and $Q$ are unimodular matrices, and, since we have only two angles in this case,  $K = \diag (k_1,k_2)$. Letting $\eta = Q x$, then reduces the unperturbed angle dynamics on the resonant torus, $\Omega(z^*) = \Omega^*$ of \Eq{Perturbed} to
\[
	\eta' = \eta + Q\Omega^* = \eta + \hat \omega \;,
\]
where $\hat \omega_i = \frac{\hat n_i}{k_i}$.
Note that this map has period $k_2$, since, as discussed in \App{Integrable}, $k_1$ divides $k_2$.

To find the map near a double resonance, we can now follow the same steps we used in \Sec{SingleResonance}: expand \Eq{Perturbed} about the integrable torus using \Eq{zScale} with exponent $p = \tfrac12$, set $\delta = 0$, and iterate the resulting system $k_2$ times. This gives
\beq{doubleResonance}
	\begin{array}{rl}
	\eta'     &= \eta + \sqrt{\eps} \alpha \zeta'  \\
	\zeta' &= \zeta + \sqrt{\eps} \bar g(\eta)
	\end{array} + \cO(\eps) \;,
\eeq
where $\bar g(\eta)$ is obtained by iterating $k_2$ times along the unperturbed torus map
\[
	\bar g(\eta) = \sum_{j=0}^{k_2-1} g(\eta + j\hat \omega)
\]
with $g(\eta)= Z(Q^{-1}\eta, z^*;0)$. Here we assume that the twist vector, $\alpha = k_2QD\Omega(z^*)$, is nonzero. This map is
a near-identity map that is, to lowest order, the time $\sqrt{\eps}$
map of the flow of the incompressible vector field
\beq{qpPend}
	\begin{array}{rl}
		\dot{\eta}     &= \alpha \zeta \;,\\
		\dot{\zeta} &=  \bar g(\eta) \;.
	\end{array}
\eeq
Note that this system still has a three-dimensional phase space, and that the ``force" $\bar g$ is a periodic function of two angles; however, these move along a fixed line on the torus. Since $\alpha$ is generically irrational this line generically has irrational slope, and the system \Eq{qpPend} is a  ``quasi-periodic pendulum". Indeed, introducing the lifted coordinate $s \in \bR$ by $\eta = \alpha s$, then
this system has the Hamiltonian
\[
	H(\eta,\zeta) = \tfrac12 \zeta^2 + V(s) \;,
\]
where $V$ is a quasi-periodic function of $s$ with derivative
$DV(s) = -\bar g(\alpha_1s, \alpha_2 s)$.  This model also occurs as a normal form for symplectic maps near a twistless singularity in the frequency map, see \cite{DIM05}.

\section{Conclusion}

We have studied resonances in near-integrable one-action, volume-preserving maps. 
Near a rank-one resonance, one component of the frequency is rational (in an appropriate basis), and when the remaining frequency components satisfy a Diophantine condition, there exists a volume-preserving coordinate transformation that decouples the fast angles from the slow angle and slow action to arbitrary order in a small parameter that represents the deviation from exact resonance. Truncation of the resulting map gives rise to a skew-product form in which the slow dynamics is given by a decoupled, area-preserving map; at lowest order, the result is \Eq{twoDAveraged}.  This map describes the dynamics of the full system on a finite, but long time-scale, as shown by \Thm{Averaging} and \Cor{VPSlow}.

There are two typical cases. If the image of the frequency map crosses the resonance transversely, then the slow map is essentially Chirikov's standard twist map \Eq{stdMap}. On the other hand, if $\Omega(z)$ has a quadratic tangency with the resonance, then the slow map is essentially the standard, nontwist map \Eq{Nontwist}.

These two cases both arise naturally from the three-dimensional map \Eq{stdTangency}, which we propose to be the volume-preserving analogue of Chirikov's standard, area-preserving map. Indeed, expanding a general one-action, volume-preserving map, near a rank-one resonance gives rise to this standard-form, \Eq{nearResonance}. Moreover, the standard map \Eq{stdTangency} itself yields this form when it is expanded about any rank-one resonance---it is recapitulated by the blow-up transformation.

The averaging results show that, near resonance, the dynamics of the three-dimensional standard map is surprisingly similar to that of the projected, two-dimensional area-preserving maps. In particular we showed that Chirikov's resonance overlap criterion can be used to estimate parameters at which invariant tori breakup near a transverse resonance crossing. 

Of course, for times longer than the $\cO(\kappa^{-1})$ of \Thm{Averaging} (and in particular when resonance overlap is significant), our simple averaging theory no longer applies and the dynamics must be treated as a intrinsically three-dimensional. For example, an understanding of the break-up of rotational tori requires a fully three-dimensional treatment \cite{Meiss11a}.

Recall that in the symplectic setting, vanishing twist is related to local non-invertibility of the frequency ratio map \cite{DIM05}. Such singularities could also be considered for the volume-preserving case.
These singularities might be global---when the frequency ratio map is not one-to-one, 
or local---when the map is not smooth at some point. An example is the development of a cusp singularity, modeled by the unfolded cusp normal form:
\begin{align*}
	\Omega(z) = (\omega_0 + z^2,  \delta z + z^3 ) \;.
\end{align*}
We plan to study such cusp singularities in a future paper.

%
%
The results outlined above hold near a rank-one resonance. By contrast, a very different reduced map \Eq{doubleResonance} is found near a rank-two (or double) resonance. It is, however, again a version of the standard map, but in this case has a quasiperiodic force.

\appendix

\section{Resonant, Integrable Torus Maps}\label{app:Integrable}

Consider the map 
\beq{torusMap}
	\theta' = \theta + \omega \mod 1
\eeq
on $\bT^d$. Suppose that $\omega \in \bT^d$ has a rank-$r$ resonance, that is the resonance module \Eq{module} has dimension $r = \dim{\cL(\omega)}$, where $0 \le r \le d$. Let $m_i \in \bZ^d, \ldots i = 1 \ldots r$ be a basis for $\cL(\omega)$, that is, a set of independent vectors so that
\[
	\cL(\omega) = \spn_\bZ(m_1,\ldots, m_r) \;.
\]
The module $\cL$ is a sublattice of $\bZ^d$. It is said to be \emph{primitive} if
\[
 \cL = \spn_\bR(m_1, \ldots, m_r) \cap \bZ^d \;,
\]
that is, every integer point in the hyperplane spanned by the vectors $m_i$ is a point in $\cL$.
Let $M$ be the $r \times d$ dimensional matrix
\[
	M^T = ( m_1, m_2, \ldots , m_r ) \;.
\]
By assumption $\rank(M) = r$. 

Following Lochak and Meunier \cite{Lochak88} and the standard normal form construction for matrices \cite [Chap. XI.7]{MacLane93}, 
there exist unimodular matrices $P$ and $Q$ such that
\beq{diagonalization}
	M = PKQ \;.
\eeq
Here $K$ is the $r \times d$ diagonal matrix
\[
	K = \diag(k_1,\ldots,k_r) \;,
\]
where $k_i \in \bN$ and $k_i$ divides $k_j$ for $i < j$. The integers $k_i$ are the \emph{invariants} of $\cL$. They determine when the lattice is primitive:

\begin{lem} A module is primitive if and only if all its invariants are one, $k_i = 1$.
\end{lem}
 
Defining the $r$-dimensional vector $n$ by $n_i = \omega \cdot m_i$, $i = 1,\ldots, r$ then
\[
	n = M\omega = PKQ\omega \;.
\]
If we define $\hat \omega = Q\omega$, then we have
\[
	K\hat \omega = P^{-1} n = \hat n \;;
\]
thus $\hat\omega_i = \frac{\hat n_i}{k_i}$ for $i = 1, \ldots, r$, while $\hat \omega_i$ for $i>r$ are incommensurate.

We use the $d \times d$ matrix $Q$ to construct a new set of angles on $\bT^d$
\[
		\psi = Q\theta 
\]
so that the map \Eq{torusMap} transforms to $\psi' = \psi + \hat \omega$, or in components
\begin{align*}
	\psi_i' &= \psi_i + \frac{\hat n_i}{k_i} \;, \quad i = 1,\ldots, r \;, \\
	\psi_k' &= \psi_k + \hat\omega_k \;, \quad k = r+1, \ldots, d \;.
\end{align*}
The orbits of the first $r$ phases are periodic with period
\beq{resonantPeriod}
	k_r = \lcm(k_1,\ldots, k_r) \;,
\eeq
and since by assumption, there are no additional resonance relations, the orbits of the last $d-r$ components are dense on $\bT^{d-r}$. Thus we may conclude that 

\begin{lem} When $\omega$ is $r$-resonant, i.e., $r = \dim{\cL(\omega)}$, the orbit of each initial condition of the integrable torus map \Eq{torusMap} is dense on a family of $k_r$, \Eq{resonantPeriod}, disjoint, $d-r$-dimensional tori that are mapped into each other by iteration.
\end{lem}

\begin{example}
For $d=3$, the frequency $\omega_a = (\sqrt{5},2\sqrt{5} + \tfrac12,\sqrt{2})$ is $1$-resonant: in order that $m\cdot \omega = n$, then $m_3 = 0$, $m_1+2m_2 = 0$, and $m_2 \in 2 \bZ$. Thus the module for this resonance is
\[
	\cL(\omega_a) = \{(-4j,2j,0): j \in  \bZ\} \subset  \bZ^3 ,
\]
which has dimension one with a basis vector $m = (-4,2,0)$. 
This resonance is not primitive since the components of the minimal basis vector $m$ are not coprime. Indeed the line $m t$ contains the integer vector $(-2,1,0)$ that is not in $\cL(\omega_a)$.

For this case, the normalization \Eq{diagonalization} results in
\[
	M = \begin{pmatrix} -4 & 2 & 0 \end{pmatrix} = 
	(1) 
	 \begin{pmatrix} 2 & 0 & 0 \end{pmatrix} 
	 \begin{pmatrix} 	
					-2 & 1 & 0 \\
				    1 & 0 & 0 \\
				    0 & 0 & 1
	\end{pmatrix} = PKQ	 \;.
\]
Thus $\hat n = P n = 1$, and
\[
	\hat \omega = Q\omega = 
	           \begin{pmatrix} 
	             \tfrac12 \\ \sqrt{5} \\ \sqrt{2}
	           \end{pmatrix} \;,
\]
so that we have $K\hat \omega = \hat n$ as required.
Note that the transformation $Q$ is orientation reversing, $\det{Q} = -1$; however we could fix this by exchanging the last two columns.
\end{example}

\begin{example}
	The frequency $\omega_b = (\sqrt{5}, \sqrt{5},2)$ has rank $2$ with module
\[
	\cL(\omega_b) = \{j_1(1,-1,0) + j_2(0,0,1): j_i\in  \bZ\} \;.
\]
Here
\[
	M = \begin{pmatrix} 1 & -1 & 0 \\ 0 & 0  & 1 \end{pmatrix}
\]
is a minimal basis for $\cL$.
This module is primitive.
\end{example}

\begin{example}
The frequency $\omega_c = (2\sqrt{2}+\tfrac12, -\sqrt{2}, \tfrac12)$ is $2$-resonant. A basis is
\[
	M = \begin{pmatrix} 1 & 2 & 1 \\ 1 & 2 & 3 \end{pmatrix} \;,
\]
with $n = (1,2)^T$.

The normalization \Eq{diagonalization} is
\[
	P = \begin{pmatrix} 1 & 0 \\ 1 & 1 \end{pmatrix}\;, \quad
	K = \begin{pmatrix} 1 & 0 & 0 \\ 0 & 2 & 0 \end{pmatrix} \;, \quad 
	Q = \begin{pmatrix} 1 & 2 & 1 \\
						0 & 0 & 1 \\
						0 & 1 & 0
		\end{pmatrix} \;,
\]
\end{example}
so that
\[
	\hat\omega = Q \omega 
		= \begin{pmatrix}  1 \\ \tfrac12 \\ -\sqrt{2} \end{pmatrix} \;, \quad
	\hat n = P^{-1} n 
		= \begin{pmatrix} 1 \\ 1 \end{pmatrix} \;.
\] 
The orbits of this torus map lie on two disjoint circles.

\begin{example}
	If $\omega \in  \bR^3$ has rank three, then its components are rational. Indeed, if $m_i \cdot \omega = n_i$ for three independent $m_i$, then  $\omega = M^{-1} n$, where the $3\times 3$ matrix $M$ is nonsingular by assumption. Note, however, that this does not imply that the module is primitive.  Instead, the torus decomposes into period $k_r$ orbits.
\end{example}

\newcommand{\etalchar}[1]{$^{#1}$}


\begin{thebibliography}{RKKCA93}

\bibitem[Arg00]{Argarwal00}
R.P. Argarwal.
\newblock {\em Difference Equations and Inequalities: Theory, Methods and
  Applications}, volume 228 of {\em Monographs and Textbooks in Pure and
  Applied Mathematics}.
\newblock Marcel Dekker Inc., New York, 2000.

\bibitem[ATPM06]{Anderson06}
P.D. Anderson, D.J. Ternet, G.~W.~M. Peters, and H.~E.~H. Meijer.
\newblock Experimental/numerical analysis of chaotic advection in a
  three-dimensional cavity flow.
\newblock {\em Int. Polymer Processing}, 04:412--420, 2006.

\bibitem[BDS98]{Bazzani98}
A.~Bazzani and A.~Di~Sebastiano.
\newblock Perturbation theory for volume-preserving maps: application to the
  magnetic field lines in plasma physics.
\newblock In {\em Analysis and modelling of discrete dynamical systems},
  volume~1 of {\em Adv. Discrete Math. Appl.}, pages 283--300. Gordon and
  Breach, Amsterdam, 1998.

\bibitem[Br{\"a}09]{Brannstrom09}
N.~Br{\"a}nnstr{\"o}m.
\newblock Averaging in weakly coupled discrete dynamical systems.
\newblock {\em J. Non. Math. Phys.}, 16(4):465--487, 2009.

\bibitem[BSV08a]{Broer08a}
H.~Broer, C.~Sim{\'o}, and R.~Vitolo.
\newblock The {H}opf-saddle-node bifurcation for fixed points of
  {3D}-diffeomorphisms: the {A}rnold resonance web.
\newblock {\em Bull. Belg. Math. Soc. Simon Stevin}, 15(5):769--787, 2008.

\bibitem[BSV08b]{Broer08b}
H.~W. Broer, C~Sim{\'o}, and R.~Vitolo.
\newblock {H}opf-saddle-node bifurcation for fixed points of
  {3D}-diffeomorphisms: Analysis of a resonance `bubble'.
\newblock {\em Phys. D}, 237:1773--1799, 2008.

\bibitem[CFP94]{Cartwright94}
J.H.E. Cartwright, M.~Feingold, and O.~Piro.
\newblock Passive scalars and three-dimensional {L}iouvillian maps.
\newblock {\em Physica D}, 76:22--23, 1994.

\bibitem[CFP96]{Cartwright96}
J.H.E. Cartwright, M.~Feingold, and O.~Piro.
\newblock Chaotic advection in three dimensional unsteady incompressible
  laminar flow.
\newblock {\em J. Fluid Mech.}, 316:259--284, 1996.

\bibitem[CFP99]{Cartwright99}
J.H.E. Cartwright, M.~Feingold, and O.~Piro.
\newblock An introduction to chaotic advection.
\newblock In H.~Chat{\'e}, E.~Villermaux, and J.M. Chomez, editors, {\em
  Mixing: Chaos and turbulence}, pages 307--342. Kluwer, 1999.

\bibitem[CK08]{Cristadoro08}
G.~Cristadoro and R.~Ketzmerick.
\newblock Universality of algebraic decays in {H}amiltonian systems.
\newblock {\em Phys. Rev. Lett.}, 100:184101, 2008.

\bibitem[CS90]{Cheng90b}
C.-Q. Cheng and Y.-S. Sun.
\newblock Existence of invariant tori in three-dimensional measure-preserving
  mappings.
\newblock {\em Celestial Mech. Dynam. Astronom.}, 47(3):275--292, 1990.

\bibitem[DEV04]{Dumas04}
H.S. Dumas, J.A. Ellison, and M.~Vogt.
\newblock First-order averaging principles for maps with applications to
  accelerator beam dynamics.
\newblock {\em {SIAM} J. Appl. Dyn. Sys.}, 3(4):409--431, 2004.

\bibitem[DIM06]{DIM05}
H.~R. Dullin, A.~V. Ivanov, and J.~D. Meiss.
\newblock Normal forms for {4D} symplectic maps with twist singularities.
\newblock {\em Physica D}, 215:175--190, 2006.

\bibitem[DM09]{Dullin09a}
H.R. Dullin and J.D. Meiss.
\newblock Quadratic volume-preserving maps: Invariant circles and bifurcations.
\newblock {\em {SIAM} J. Appl. Dyn. Sys.}, 8(1):76--128, 2009.

\bibitem[DMS00]{DMS98b}
H.~R. Dullin, J.~D. Meiss, and D.~G. Sterling.
\newblock Generic twistless bifurcations.
\newblock {\em Nonlinearity}, 13:203--224, 2000.

\bibitem[GR07]{GrindrodReardon}
P.~Grindrod and F.~Reardon.
\newblock On the interactions between invariant tubes in volume preserving
  maps.
\newblock Unpublished Report, 2007.

\bibitem[Gre93]{Greene93}
J.M. Greene.
\newblock Reconnection of vorticity lines and magnetic lines.
\newblock {\em Phys. Fluids B}, 5(7):2355--2362, 1993.

\bibitem[HH84]{HowHoh84}
J.E. Howard and S.M. Hohs.
\newblock Stochasticity and reconnection in {H}amiltonian systems.
\newblock {\em Physical Review A}, 29:418, 1984.

\bibitem[HW79]{HardyWright79}
G.H. Hardy and E.M. Wright.
\newblock {\em An introduction to the theory of numbers}.
\newblock Oxford Univ. Press, Oxford, 1979.

\bibitem[Kif03]{Kifer03}
Y.~Kifer.
\newblock Averaging in difference equations driven by dynamical systems.
\newblock {\em Asterisque}, 287:103--123, 2003.

\bibitem[LL92]{Lichtenberg92}
A.J. Lichtenberg and M.A. Lieberman.
\newblock {\em Regular and Chaotic Dynamics}, volume~38 of {\em Applied
  Mathematical Sciences}.
\newblock Springer-Verlag, New York, 2nd edition, 1992.

\bibitem[LM88]{Lochak88}
P.~Lochak and C.~Meunier.
\newblock {\em Multiphase averaging for classical systems: With applications to
  adiabatic theorems}, volume~72 of {\em Applied mathematical sciences}.
\newblock Springer-Verlag, New York, 1988.

\bibitem[LM09]{Lomeli09b}
H.~Lomel\'i and J.D. Meiss.
\newblock Resonance zones and lobe volumes for volume-preserving maps.
\newblock {\em Nonlinearity}, 22:1761--1789, 2009.

\bibitem[LS94]{Liu94}
J.~Liu and Y.-S. Sun.
\newblock Chaotic motion of comets in near-parabolic orbit: Mapping aproaches.
\newblock {\em Celestial Mech. Dynam. Astronom.}, 60:3--28, 1994.

\bibitem[MB93]{MacLane93}
S.~MacLane and G.~Birkhoff.
\newblock {\em Algebra}.
\newblock Chelsea Publishing, New York, 1993.

\bibitem[Mei92]{Meiss92}
J.D. Meiss.
\newblock Symplectic maps, variational principles, and transport.
\newblock {\em Reviews of Modern Physics}, 64(3):795--848, 1992.

\bibitem[Mei07]{Meiss07a}
J.D. Meiss.
\newblock {\em Differential Dynamical Systems}, volume~14 of {\em Mathematical
  modeling and computation}.
\newblock SiAM, Philadelphia, 2007.

\bibitem[Mei11]{Meiss11a}
J.D. Meiss.
\newblock The destruction of tori in volume-preserving maps.
\newblock {\em Communications in Nonlinear Science and Numerical Simulation},
  in press, 2011.

\bibitem[Mez01]{Mezic01}
I.~Mez{\'\i}c.
\newblock Break-up of invariant surfaces in action-angle-angle maps and flows.
\newblock {\em Phys. D}, 154(1-2):51--67, 2001.

\bibitem[MGPM99]{Meleshko99}
V.V. Meleshko, O.S. Galaktionov, G.W.M. Peters, and H.E.H. Meijer.
\newblock Three-dimensional mixing in {S}tokes flow: The partitioned pipe mixer
  problem revisited.
\newblock {\em Eur. J. Mech. B - Fluids}, 18:783--792, 1999.

\bibitem[MJM08]{Mullowney08}
P.~Mullowney, K.~Julien, and J.D. Meiss.
\newblock Chaotic advection in the {K}\:uppers-{L}ortz state.
\newblock {\em Chaos}, 18:033104, 2008.

\bibitem[MMP84]{MacKay84}
R.S. Mac{K}ay, J.D. Meiss, and I.C. Percival.
\newblock Transport in {H}amiltonian systems.
\newblock {\em Physica D}, 13:55--81, 1984.

\bibitem[MO86]{Meiss86}
J.D. Meiss and E.~Ott.
\newblock Markov tree model of transport in area preserving maps.
\newblock {\em Physica D}, 20:387--402, 1986.

\bibitem[Mos66]{Moser66}
J.K. Moser.
\newblock On the theory of quasiperiodic motions.
\newblock {\em SIAM Review}, 8:145--171, 1966.

\bibitem[P\"93]{Poshel93}
J.~P\"oschel.
\newblock Nekhoroshev estimates for quasi-convex {H}amiltonian systems.
\newblock {\em Math. Zeit.}, 213:187--216, 1993.

\bibitem[PF88]{Piro88}
O.~Piro and M.~Feingold.
\newblock Diffusion in three-dimensional {L}iouvillian maps.
\newblock {\em Phys. Rev. Lett.}, 61:1799, 1988.

\bibitem[PW94]{Perry94}
A.D. Perry and S.~Wiggins.
\newblock {KAM} tori are very sticky: rigorous lower bounds on the time to move
  away from an invariant {L}agrangian torus with linear flow.
\newblock {\em Physica D}, 71:102--121, 1994.

\bibitem[RKKCA93]{RomKedar93}
V.~Rom-Kedar, L.P. Kadanoff, E.S. Ching, and C.~Amick.
\newblock The break-up of a heteroclinic connection in a volume preserving
  mapping.
\newblock {\em Physica D}, 62(1--4):51--65, 1993.

\bibitem[RML{\etalchar{+}}03]{Rodrigo03}
A.J.S. Rodrigo, J.P.B. Mota, A.~Lefevre, J.C. Leprevost, and E.~Saatdjian.
\newblock Chaotic advection in a three-dimensional {S}tokes flow.
\newblock {\em AIChE Journal}, 49(11):2749--2758, 2003.

\bibitem[SCVH04]{Speetjens04}
M.F.M. Speetjens, H.J.H. Clercx, and G.J.F. Van~Heijst.
\newblock A numerical and experimental study on advection in three-dimensional
  {S}tokes flows.
\newblock {\em J. Fluid Mech}, 514:77--105, 2004.

\bibitem[Sim98]{Simo98}
C.~Sim{\'o}.
\newblock Invariant curves of analytic perturbed nontwist area preserving maps.
\newblock {\em Regular {\&} Chaotic Dynamics}, 3:180--195, 1998.

\bibitem[SVL01]{Sotiropoulos01}
F.~Sotiropoulos, Y.~Ventikos, and T.~C. Lackey.
\newblock Chaotic advection in three-dimensional stationary vortex-breakdown
  bubbles: \v{S}il'nikov's chaos and the devil's staircase.
\newblock {\em Journal of Fluid Mechanics}, 444:257--297, 2001.

\bibitem[SVM07]{Sanders07}
J.~A. Sanders, F.~Verhulst, and J.~Murdock.
\newblock {\em Averaging Methods in Nonlinear Dynamical Systems}, volume~59 of
  {\em Appl. Math. Sciences}.
\newblock Springer, New York, 2007.

\bibitem[SZ08]{Sun08}
Y.-S. Sun and L.-Y. Zhou.
\newblock Stickiness in three-dimensional volume preserving mappings.
\newblock {\em Celestial Mech. Dynam. Astronom.}, 103(2):119--131, 2008.

\bibitem[TH85]{Thyagaraja85}
A.~Thyagaraja and F.A. Haas.
\newblock Representation of volume-preserving maps induced by solenoidal vector
  fields.
\newblock {\em Phys. Fluids}, 28(3):1005--1007, 1985.

\bibitem[Ven09]{Venegeroles09}
R.~Venegeroles.
\newblock Universality of algebraic laws in {H}amiltonian systems.
\newblock {\em Phys. Rev. Lett.}, 102:064101, 2009.

\bibitem[WAFM05]{Wurm05}
A.~Wurm, A.~Apte, K.~Fuchss, and P.J. Morrison.
\newblock Meanders and reconnection-collision sequences in the standard
  nontwist map.
\newblock {\em Chaos}, 15:023108, 2005.

\bibitem[Xia92]{Xia92}
Z.~Xia.
\newblock Existence of invariant tori in volume-preserving diffeomorphisms.
\newblock {\em Erg Th Dyn Sys}, 12(3):621--631, 1992.

\end{thebibliography}
\end{document}